\documentclass{article}

\usepackage{arxiv}

\usepackage[utf8]{inputenc} 
\usepackage[T1]{fontenc}    
\usepackage{hyperref}       
\usepackage{url}            
\usepackage{booktabs}       
\usepackage{amsfonts}       
\usepackage{nicefrac}       
\usepackage{microtype}      
\usepackage{lipsum}		
\usepackage{graphicx}
\usepackage[numbers]{natbib}
\usepackage{doi}
\usepackage[nohyperlinks, nolist]{acronym} 
\usepackage{tabularx}
\usepackage{cleveref}
\usepackage{dirtytalk}
\usepackage[inline]{enumitem}
\usepackage{tikz}
\usepackage{graphicx}          
\usetikzlibrary{arrows.meta}
\usepackage[T1]{fontenc}
\usepackage{graphicx}
\usepackage{todonotes}
\usepackage[nolist]{acronym}
\usepackage{cleveref}
\usepackage{subcaption}
\usepackage{booktabs}
\usepackage{multirow}
\usepackage{tcolorbox}

\usepackage{lscape}
\usepackage{tikz}
\usetikzlibrary{positioning,arrows.meta,fit,backgrounds,calc,shapes.geometric}

\title{Communicating Uncertainty in AI-Supported Nurse Scheduling: Expert Co-Design and Evaluation of an Explainable AI Prototype}

\date{} 					

\author{Alice Rollwagen\thanks{Alice Rollwagen and Emily Schiller contributed equally to this research.} \\
	XITASO GmbH, Augsburg, Germany \\
    Human-Computer Interaction Group (HCIG), Faculty of Computer Science \\
    Technische Hochschule Ingolstadt, Ingolstadt, Germany \\
    \texttt{alice.rollwagen@thi.de} \\
	\And
	Emily Schiller\footnotemark[1] \\
	XITASO GmbH, Augsburg, Germany \\
    The Artificial Intelligence and Cognitive Load Research Lab \\
    University College Cork, Cork, Ireland \\
    \texttt{emily.schiller@xitaso.com} \\
    \And
	Anton Hummel \\
	XITASO GmbH, Augsburg, Germany \\
    University of Bayreuth, Bayreuth, Germany \\
    \texttt{anton.hummel@xitaso.com} \\
    \And
    Kathrin Ebertsch \\
	Augsburg University Hospital, \\ Department for Digitalization and Nursing Science, \\ Augsburg, Germany \\
    \texttt{kathrin.Ebertsch@uk-augsburg.de} \\
    \And
	Andreas Riener \\
	Human-Computer Interaction Group (HCIG), Faculty of Computer Science \\
    Technische Hochschule Ingolstadt, Ingolstadt, Germany \\
    \texttt{andreas.riener@thi.de} \\
    \And
	Jan-Philipp Steghöfer \\
	XITASO GmbH, Augsburg, Germany \\
    \texttt{jan-philipp.steghoefer@xitaso.com} \\
}

\renewcommand{\shorttitle}{Communicating Uncertainty in AI-Supported Nurse Scheduling}

\hypersetup{
pdftitle={A template for the arxiv style},
pdfsubject={q-bio.NC, q-bio.QM},
pdfauthor={David S.~Hippocampus, Elias D.~Striatum},
pdfkeywords={First keyword, Second keyword, More},
}

\begin{document}

\begin{acronym}
    \acro{HCI}{Human-Computer Interaction}
    \acro{AI}{Artificial intelligence}
    \acro{CDSS}{clinical decision support system}
    \acro{DSS}{decision support systems}
\end{acronym}
\maketitle

\begin{abstract}
Nurse scheduling is a complex optimization problem in which nurses are assigned to shifts to meet staffing demands while complying with working-time limits, qualification requirements, and other operational and regulatory constraints~\cite{NRP}. Artificial intelligence is increasingly being used to support this process by forecasting future staffing needs and providing advice on how to improve shift scheduling. Effective human-AI collaboration in nurse scheduling requires users to understand, interpret, and appropriately act on AI forecasts and advice. However, it remains insufficiently understood how healthcare professionals use transparency mechanisms like uncertainty communication and AI advice explanations when making staffing decisions.
Applying a human-centered design approach, we iteratively developed a prototype decision support system for nurse scheduling and evaluated it through a mixed-methods user study with expert users. Using thematic analysis, questionnaire data, and eye tracking, we examine how healthcare professionals interact with the system and how uncertainty communication and explanations of AI advice influence understanding, decision confidence, trust, and usability.
We found that the uncertainty communicated in AI forecasts was frequently misinterpreted and reduced decision confidence. Explanations were generally perceived positively, but did not significantly affect decision confidence or advice acceptance. We conclude with eight design implications derived from our findings, providing insights for the design of more transparent and effective AI-supported decision support tools in healthcare.
\end{abstract}


\begin{figure}
    \includegraphics[width=\linewidth]{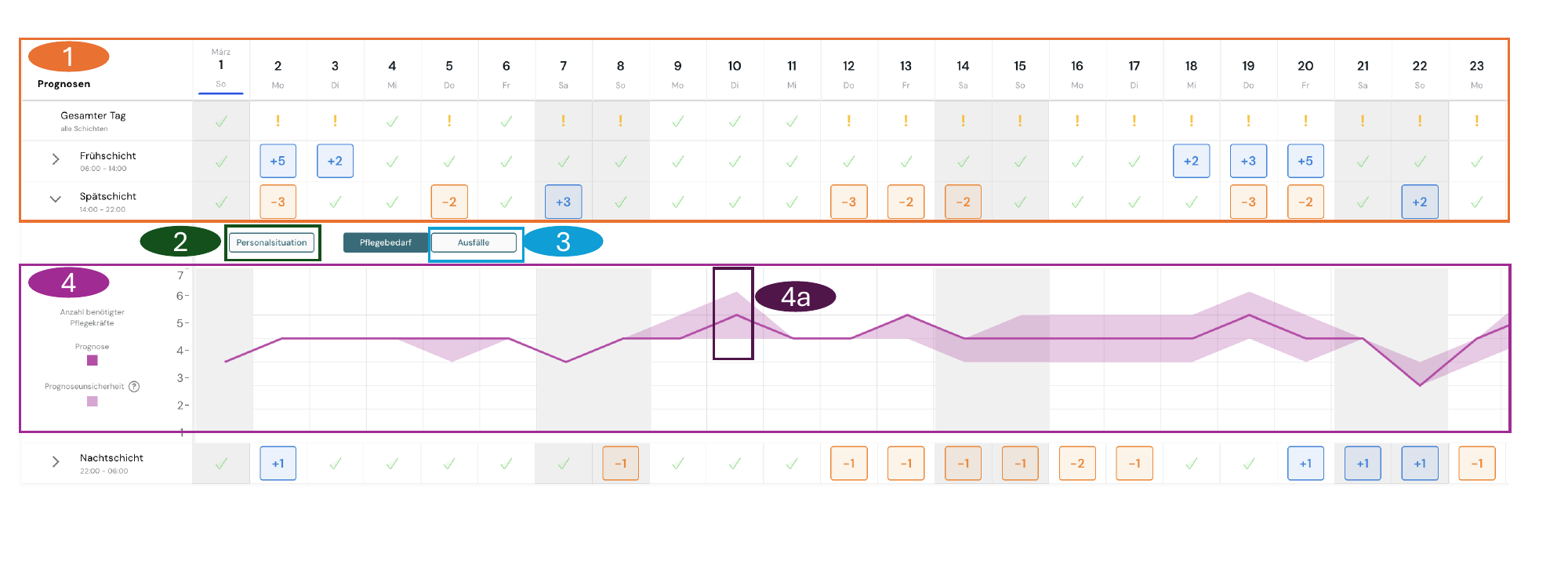}
    \caption{Visualization of the developed AI-based decision support system for nurse scheduling. The image shows the time series forecast view with forecasts of future care demand and staff absences to support the creation of shift plans. Box (1) contains an overview showing whether, based on the forecasts, action is needed on a specific day or during a shift to avoid understaffing or overstaffing. Box (2) highlights the tab that takes you to a comparison of care demand and planned capacity. Box (3) displays the tab that takes you to the forecasts of staff absences. Box (4) displays the forecasts of care demand with (4a) highlighting a day with high prediction uncertainty.}
    \label{fig:tsf}
\end{figure}

\maketitle

\section{Introduction}

Healthcare systems worldwide are facing increasing challenges due to demographic change, rising patient numbers, and persistent shortages in qualified healthcare professionals~\cite{wynendaele2021systematic}. Nurse scheduling refers to the process of assigning nurses to shifts over a planning horizon of typically four to ten weeks, while satisfying a range of requirements, including legal and contractual regulations such as minimum rest periods, minimum staffing levels, as well as qualification requirements~\cite{NRP}. This makes nurse scheduling a particularly complex and time-consuming~\cite{wynendaele2021systematic, Rönnberg2013, mohammadian2019scheduling} yet critical organizational decision-making task with high impact on nurse job satisfaction~\cite{RizanyNurseSatisfactionScheduling, mohammadian2019scheduling}, patient satisfaction~\cite{WINTER2020380}, and the overall quality of hospital care~\cite{mohammadian2019scheduling, CHEANG2003447}. AI-supported systems offer a promising avenue to address these challenges by helping planners construct schedules that better align nursing capacity with patient care demand~\cite{renggli_integrating_2025,LEUNG2022373, OLYA2022115924,sandow2024dynamic}. However, the effectiveness of decision support systems (DSS) depends not only on predictive or optimization performance but also on how users perceive and interact with AI-generated recommendations~\cite{userstudy}. If recommendations are insufficiently transparent, users may ignore useful suggestions or over-rely on them without appropriate scrutiny~\cite{renggli_integrating_2025}. Particularly in safety-critical domains such as healthcare, users remain responsible for final decisions and therefore require sufficient information to critically assess AI-generated recommendations. Human-centered AI research has emphasized fostering effective human–AI collaboration through the design of AI systems that promote appropriate reliance and transparency \cite{cao_designing_2024, Voessing2022Designing,Gomez2023Human-AI}.

Two complementary transparency mechanisms have emerged as promising directions for supporting users of AI-supported DSS. The first is explainable AI (XAI), which aims to increase the transparency of AI advice by communicating why a particular piece of advice has been made. Explanations can help users assess the plausibility of AI advice and increase confidence in AI-supported decisions~\cite{userstudy}. 
The second mechanism is the communication of prediction uncertainty. Predictions generated by AI systems inherently contain uncertainty arising from incomplete information, changing conditions, or variability in the data. Nevertheless, many DSS omit uncertainty information in their interfaces, potentially encouraging users to overestimate the system's reliability. Communicating uncertainty has the potential to support users in interpreting AI outputs more realistically, enabling more informed decision-making and better trust calibration. 

However, XAI and uncertainty communication come with their own pitfalls~\cite{ehsanExplainabilityPitfallsDark2024} and limitations~\cite{martensBewareExplanationsAI2025}. For instance, explanations are not human-understandable~\cite{ghassemiFalseHopeCurrent2021}, fail to address trust or bias mitigation~\cite{ghassemiFalseHopeCurrent2021,deckCriticalSurveyFairness2024}, and ultimately can lead to overtrust or mistrust~\cite{schemmerAppropriateRelianceAI2023}. Similarly, uncertainty information introduces additional cognitive demands and may reduce usability if presented inappropriately~\cite{bhatt2021uncertainty}. Existing research has reported mixed findings regarding how uncertainty visualizations affect user understanding, trust, and decision quality~\cite{prabhudesai2023understanding,cao_designing_2024,reyes_trusting_2025,schafer_ai_2025}.

Although both mechanisms are widely recognized as important design principles for trustworthy AI, they have largely been investigated separately. Further, empirical evidence from realistic healthcare planning user studies involving domain experts remains limited. In particular, there is little understanding of how these mechanisms impact the acceptance and use of AI-supported nurse scheduling.

To address this gap, we conduct a mixed-methods user study including eye tracking, evaluating an AI-supported nurse scheduling prototype we developed within a research project for clinical workforce planning. The study investigates two complementary interface features. First, we examine how well visualizations of predictive uncertainty for time series forecasts of care demand and staff absences are understood and how they influence decision confidence, task performance, and usability. Second, we evaluate how explanations accompanying AI-generated nurse scheduling advice influence acceptance, trust, decision confidence, and user experience. The study aims to answer the following research questions:

\begin{description}
    \item[RQ1:] How well is the uncertainty information understood, and to what extent do brief explanations of how to interpret the uncertainty information improve understanding and interpretation? %

    \item[RQ2:] How does uncertainty visualization influence users’ confidence in decision-making?

    \item[RQ3:] How do explanations of AI recommendations influence users’ acceptance and their confidence in decision-making?

\end{description}

\subsubsection*{Contribution Statement:}
Our contributions to Human-Computer Interaction (HCI)~\cite{lazar2017research}[ch.~1.2] are:
\begin{enumerate}
\item \textbf{Design and evaluation of an AI-supported nurse scheduling prototype:} We designed and evaluated an interactive AI-supported nurse scheduling prototype that combines time series forecasts, uncertainty communication, visualizations of historical forecast performance, and explainable AI advice to investigate how different forms of AI information support shift-planning decisions.
\item \textbf{Empirical study of uncertainty communication and explainable AI in shift planning:} We conducted a mixed-methods user study with nurses and shift planners, showing that prediction intervals for uncertainty communication were often misunderstood, while AI explanations improved perceived utility, but did not significantly affect advice acceptance or decision confidence.
\item \textbf{Design implications for AI-supported nurse scheduling:} Based on these findings, we derived design implications to support comprehensible uncertainty communication, concise explanations, calibrated trust, human oversight, and AI-supported socially aware nurse scheduling.
\end{enumerate}

By combining quantitative measures with qualitative insights from healthcare professionals, our work contributes empirical evidence on the design of intelligent user interfaces for human-centered AI-supported clinical decision support.

\section{Related Work}

\subsection{Clinical Decision Support for Nurse Scheduling}

Nursing workforce planning involves determining and deploying the number and qualifications of nurses required to ensure safe patient care~\cite{griffiths2020nursing}. This process is shaped by external factors\,--\,such as legal frameworks\,--\, and internal operational demands. Regulatory care demand assessment frameworks like Germany's \textit{Pflegepersonalregelung 2.0} (PPR)~\cite{schmedders2023pflegepersonalvorgaben} or, more generally, patient-to-nurse ratios, specify the minimum number of required nursing staff per shift. 
In most hospital wards, shift plans are prepared on a monthly basis, usually four to ten weeks in advance starting with coverage requirements such as staff qualifications and staffing demand across early, late, and night shifts~\cite{ronnberg2012automatic}. This process, called \textit{nurse scheduling}, is still often carried out manually by ward managers, usually with the support of planning tools. However, manual planning is particularly challenging because it must take into account staff preferences, internal requirements and patient volume~\cite{LEUNG2022373}, among others. Notably, tensions between scheduling for patient care needs and optimizing staff needs have been identified~\cite{stimpfel2025scheduling}. Operational shift planning still relies heavily on expert knowledge and current indicators, while future developments and anticipated changes in care demand are often neglected, leading to overly uniform staffing estimates. AI forecasts can equip shift planners with valuable foresight into future staffing demands~\cite{11559628}. Several studies have therefore examined the prediction of care demand, for example on the basis of patient admissions or bed occupancy~\cite{caldas2022temporal,zhang2022forecasting,TUOMINEN2023}, or staff absences~\cite{schiller2024no}. 

A small number of studies have evaluated AI-supported scheduling tools in real-world settings. 
\citet{userstudy} developed an AI-supported scheduling DSS that automatically generates complete shift plans containing a workload prediction component with SHAP-based explanations~\cite{lundberg2017unified}. Deployed across 156 nurses, the system achieved substantial reductions in scheduling time and errors, and nurse engagement with the explanations was positively associated with satisfaction, underscoring the importance of transparency for adoption. \citet{renggli2025integrating} conducted focus group interviews with 21 nurses and supervisors and found that while a majority saw potential for improved efficiency and fairness, 38\% expressed concerns about AI reliability and loss of human oversight. This tension between efficiency gains and the desire for human control motivates the design of DSS that augment rather than replace human judgment, which was a core principle in the development of the system evaluated in this study. Moreover, we build upon the concerns about AI reliability and transparency by incorporating AI uncertainty and explainable AI. To the best of our knowledge, no study has examined these aspects in a real-world nurse scheduling context through a dedicated user study.

\subsection{Transparency Mechanisms for Human-AI Decision-Making}

AI-supported clinical DSS often suffer from the opacity of the underlying AI model, a challenge commonly referred to as the ``black-box'' problem~\cite{rudinStopExplainingBlack2019}. This lack of transparency can give rise to well-documented problems in human-computer interaction, such as over-reliance or under-reliance on the system~\cite{spitzerImperfectionsXAIPhenomena2025}, as well as anchoring and confirmation bias~\cite{wangDesigningTheoryDrivenUserCentric2019}. To address these challenges, different transparency mechanisms---in particular explainable AI (see~\Cref{sec:xai}) and uncertainty communication (see~\Cref{sec:uq})---have emerged as helpful tools for achieving trust~\cite{zhanExplainableAITrust2025} and appropriate reliance~\cite{schemmerAppropriateRelianceAI2023} and thereby improving human-AI decision-making.

\subsubsection{Explainable AI}\label{sec:xai}

Explainable AI (XAI) has emerged as a research field to increase transparency of AI models by providing explanations that simplify AI outputs for human understanding~\cite{LONGO2024102301}. Although XAI is being used more often in applications and software for end users, human-centered evaluations remain relatively limited and rarely incorporate insights from cognitive or social sciences~\cite{Rong_2024,millerExplanationArtificialIntelligence2019}. For this reason, human-centered explainable AI (HCXAI) has emerged as an approach that places users and their desiderata at the center of the design process~\cite{Ehsan_2020, Leersum_2025,Ridley_2024}. It extends beyond the technical generation of explanations by considering human values and desiderata~\cite{langerWhatWeWant2021}, interpersonal dynamics, and the broader social context~\cite{millerExplanationArtificialIntelligence2019} in which AI systems are used. This also includes who needs an explanation, why it is needed and in which context it is used, because different social groups interpret explainability in different ways and use it for different purposes~\cite{Dhanorkar_2021, Ehsan_2020}. Consequently, HCXAI emphasizes the joint examination of technical explainability and human factors, including how users perceive, understand, and evaluate AI-generated explanations.

\subsubsection{Communicating Uncertainty in AI Systems}\label{sec:uq}

Uncertainty is an inherent property of machine learning predictions and is commonly decomposed into two types~\cite{hullermeier2021aleatoric}. \textit{Aleatoric uncertainty} reflects irreducible variability in the data arising from stochasticity or ambiguity. \textit{Epistemic uncertainty}, or model uncertainty, stems from limitations in the model's knowledge, for instance, due to insufficient training data. \textit{Predictive uncertainty} (or \textit{prediction uncertainty}) integrates both sources. Common approaches for quantifying predictive uncertainty include Deep Ensembles~\cite{lakshminarayanan2017simple} and Monte Carlo Dropout~\cite{gal2016dropout}, both of which produce a distribution of predictions for a given input from which statistical uncertainty estimates---such as variance or prediction intervals---can be derived.

Communicating uncertainty to end-users remains a significant challenge, owing to variability in statistical literacy and the inherent complexity of uncertainty representations such as variances, probabilities, and prediction intervals~\cite{bhatt2021uncertainty}. Several recent studies have examined how uncertainty information shapes user behavior in AI-supported decision-making~\cite{reyes_trusting_2025,schafer_ai_2025,cao_designing_2024,mcgrath2023when}. \citet{cao_designing_2024} investigated the effect of uncertainty presentation format on appropriate reliance in a skin cancer screening task. They found that expressing model uncertainty in a frequency format enabled users to adjust their reliance more appropriately and reduced the influence of confirmation bias. Their results further indicated that age and prior familiarity with statistics modulated the quality of human-AI collaborative decisions. \citet{nourani_heart_2026} documented the design of a clinical decision support interface for intensive care unit risk prediction, with a particular focus on how uncertainty arising from missing data can be effectively visualized. \citet{prabhudesai2023understanding} conducted a qualitative study in which 17 lay decision-makers interacted with three decision support conditions: an interactive visualization, an ML-based system providing predictions without uncertainty information, and the same system augmented with uncertainty information. Their findings suggest that exposure to uncertainty information prompted participants to engage in more deliberate, analytical reasoning, which in turn reduced over-reliance on the ML system.

\subsection{Research Gap}

The preceding review reveals two compounding gaps in the existing literature. First, while XAI and uncertainty communication have each been recognized as important transparency mechanisms for AI-supported DSS, empirical user studies evaluating their respective effects on understanding, trust, and decision-making in realistic nurse scheduling settings with domain experts remain scarce. While the work of \citet{prabhudesai2023understanding} on uncertainty communication is most closely related to our study, their evaluation was conducted with lay users on a hypothetical house price prediction task, leaving open the question of how uncertainty communication affects expert users in real-world professional settings. Moreover, the understandability of uncertainty visualizations themselves, as opposed to their downstream effects on decision-making, has received comparatively little attention in the existing literature. Regarding XAI in a nurse scheduling context, we identified one study that evaluated explanations of workload prediction~\cite{userstudy}. However, their evaluated system differs substantially from ours: rather than generating complete shift plans autonomously, our system provides advice on which shifts to change, preserving planner control and judgment and placing greater emphasis on human autonomy and oversight.

\paragraph{Implications for this study.} These gaps motivate a mixed-methods user study with healthcare professionals, evaluating an AI-supported nurse scheduling prototype that incorporates both uncertainty visualization and XAI explanations as separate interface features. The XAI component targets shift plan advice, providing explanations for suggested shift changes while leaving final decisions to the planner. Using complementary quantitative and qualitative methods including eye tracking, structured questionnaires, and thematic analysis of think-aloud data, we investigate the effect of XAI and uncertainty communication on understanding, decision confidence, trust, and usability.

\section{Prototype Implementation}
\label{tool}

Applying a human-centered design approach based on the principles of ISO 9241-210:2019 \cite{iso9241-210}, we iteratively developed a web-based decision support prototype integrating AI forecasts and optimization-based scheduling advice as part of a larger research project on AI-supported nurse scheduling. The prototype was evaluated with domain experts and practicing shift planners and is fully functional, although it contains partially mocked data, as detailed below.

\subsection{Time Series Forecasts}
\label{sec:forecasts}

The first functionality provides time series forecasts of (a) future care demand and (b) unplanned staff absences for a hospital ward over the next 30 days, informing shift planners of shifts with potentially inadequate staffing. The two targets are defined as follows:

\begin{description}
    \item[Care demand] is operationalized via patient acuity scores under the German PPR regulation, quantifying individual care requirements in minutes per day. Total care demand per time step is the sum of acuity scores across all ward patients.

    \item[Staff absences] are predicted from historical records of \textit{unplanned absences} (e.g., sick leave), as distinct from planned absences such as scheduled vacation.

\end{description}

Both targets are forecast as hourly time series over a horizon of 720 hours (30 days) using a TSMixer model~\cite{chen2023tsmixer}, a deep learning architecture designed for long-term time series forecasting. Hourly outputs are aggregated into the three most prevalent shift types at the study site (early, late, night; each 8 hours), following a co-design session with an experienced shift planner. Care demand forecasts, expressed in PPR minutes, were converted to nursing staff equivalents assuming 480 minutes of capacity per nurse per shift.
As the source ward employs more nurses than typical study wards, forecast outputs were scaled down proportionally to represent a smaller, anonymized ward. Prediction uncertainty was communicated via 95\% prediction intervals, which were constructed manually to ensure relevance to study tasks.

Forecasts are displayed below the shift plan (see~\Cref{fig:tsf}). The interface highlights days requiring action due to over- or understaffing (\Cref{fig:tsf} (1)), and provides three additional views per shift: a combined overlay of both forecasts with current staffing levels (\Cref{fig:tsf} (2)), and dedicated views for staff absences (\Cref{fig:tsf} (3)) and care demand (\Cref{fig:tsf} (4)). Planners can toggle between shift types within each view.

\subsection{AI Advice on Nurse Scheduling}
\label{sec:optimization}

The second functionality provides data-driven advice for improving an existing shift plan by proposing targeted modifications that better align it with forecast care demand and staff absences, and regulatory requirements.

\begin{figure}[t]
    \centering
    \includegraphics[width=\linewidth]{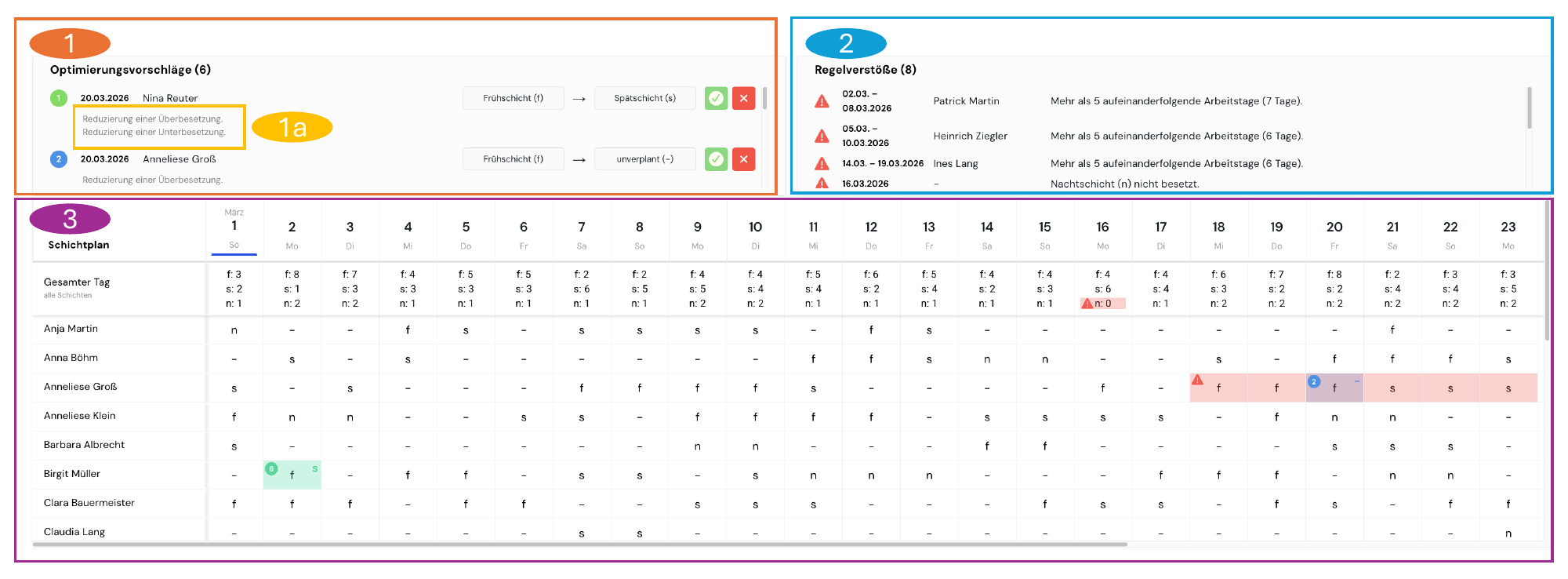}
    \caption{Visualization of the AI advice view. (1) shows the AI advice; (1a) shows an explanation for the first suggestion. (2) lists the regulatory violations. (3) displays the current shift schedule, in which the advice and regulatory violations are highlighted in the corresponding shifts. All names are fictitious. The abbreviations f, s, and n stand for early, late, and night shifts, respectively, and the top row shows the number of nursing staff assigned to each shift on a given day.}
    \label{fig:opti}
\end{figure}

AI advice is generated by a rule-based heuristic that considers the current shift plan, predicted care demand and staff absences, and regulatory scheduling constraints (e.g., no more than five consecutive working days). The heuristic minimizes a penalty function over staffing deviations and constraint violations by iteratively proposing single-shift changes; for example, reassigning a nurse from an early shift to a day off when that shift is forecast to be overstaffed. The underlying shift plan was adapted from a real ward schedule: non-standard shift types were reassigned to early, late, or night shifts, and the plan was reduced in scale to a ward of 22 nurses with fictitious names. Forecast and shift plan data are fully synchronized to ensure internal consistency.

Advice is presented as an enumerated list and highlighted within the shift plan view, enabling planners to assess the advice in context (see~\Cref{fig:opti}). Each advice item is accompanied by a concise natural language explanation derived from the penalty component most reduced by the proposed change (e.g., ``reduction of overstaffing''). Planners retain full agency since each advice item can be individually accepted or declined. Existing regulatory violations are also listed and highlighted.

\section{Study Design and Execution}
\label{task_questions}

We conducted a within-subject study comparing multiple variants of the AI-supported nurse scheduling prototype from ~\Cref{tool}. The within-subject design allowed us to compare participants' interactions and evaluations across the different design variants. During the interaction, participants wore eye tracking glasses to record their visual attention and were asked to think aloud. 
Before the experimental tasks, participants watched a video tutorial introducing the basic functionalities and interaction principles of the prototype. To avoid bias, the video included placeholder information for design elements that varied depending on the design variant.

The study comprised two blocks focusing on different aspects of AI-supported decision-making: (1) Uncertainty communication in time series forecasts, which examined how well uncertainty information was understood (RQ1) and how uncertainty visualizations influenced users' decision confidence (RQ2); and (2) AI advice for shift plan improvements, which examined how explanations of AI advice influenced user acceptance and confidence in decision-making (RQ3). 

\begin{figure}[t]
\centering

\begin{subfigure}{0.8\textwidth}

\centering
\includegraphics[width=\linewidth]{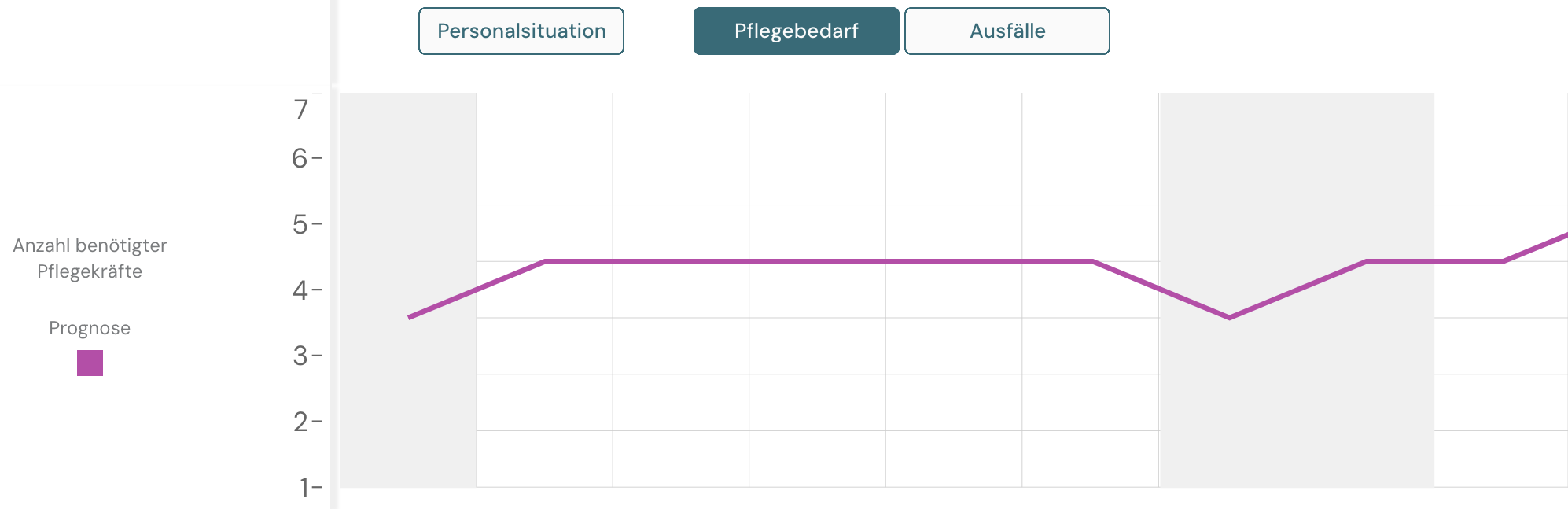}
\caption{TSF w/o UV: Time series forecasts of care demand and staff absences without visualization of predictive uncertainty.}
\end{subfigure}

\vspace{0.5cm}

\begin{subfigure}{0.8\textwidth}
\centering
\includegraphics[width=\linewidth]{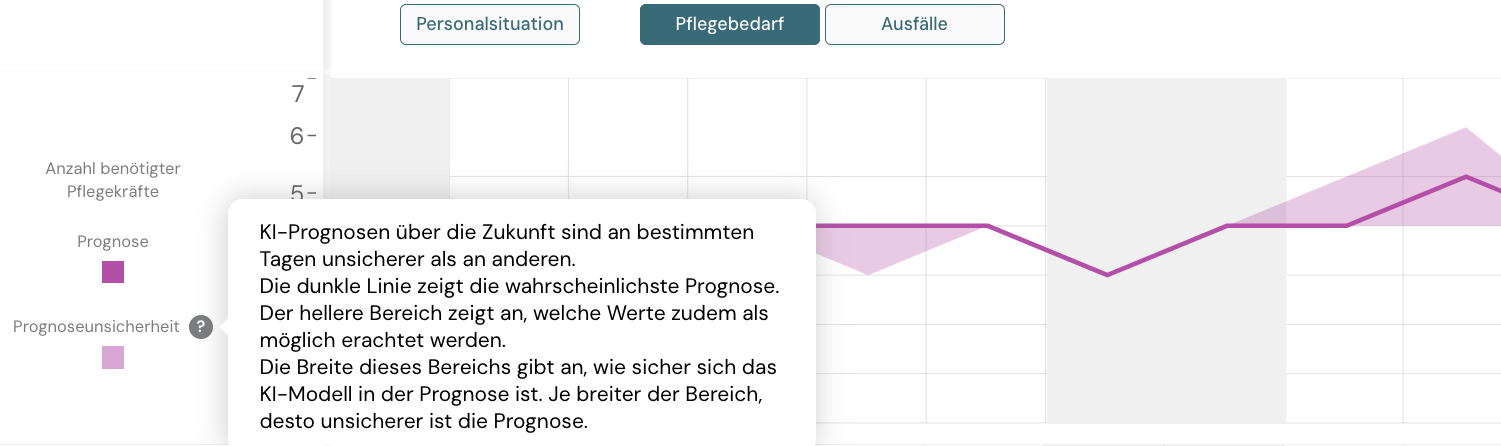}
\caption{TSF w/ UV: Time series forecasts of care demand and staff absences with visualizations of predictive uncertainty and info button with a system explanation that describes how to interpret the uncertainty visualization. The text translates to: ``\textit{AI forecasts about the future are more uncertain on some days than on others. The dark line shows the most likely forecast. The lighter area indicates the values that are also considered possible. The width of this area indicates how certain the AI model is in its forecast. The wider the area, the more uncertain the forecast.}''}
\end{subfigure}

\vspace{0.5cm}

\begin{subfigure}{0.8\textwidth}
\centering
\includegraphics[width=\linewidth]{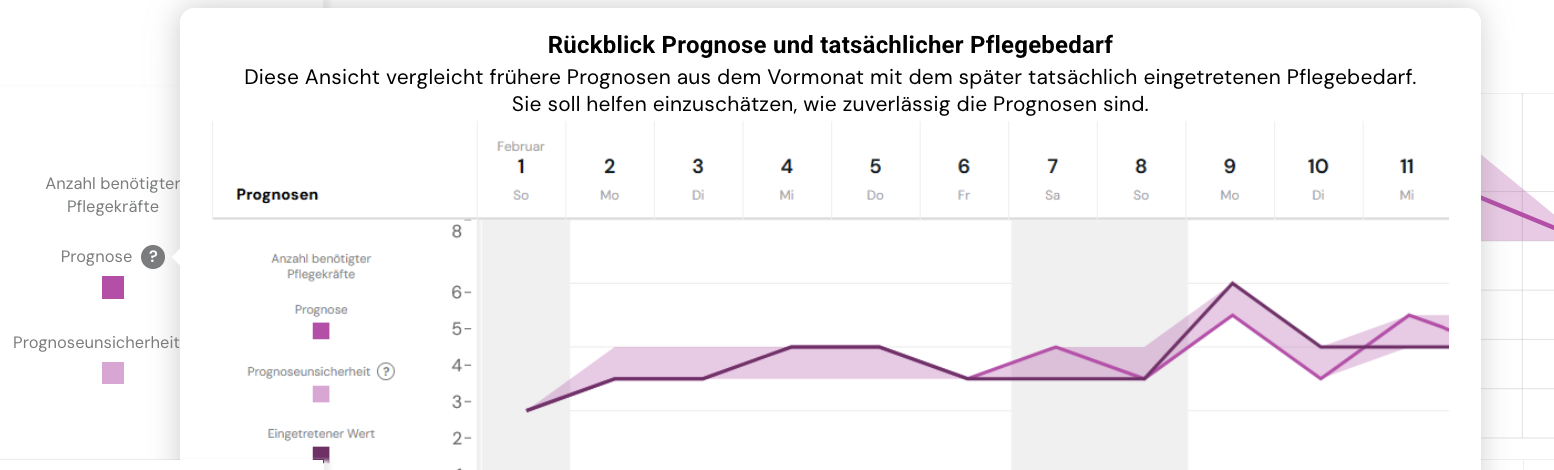}
\caption{TSF w/ UV + PH: This design variant extends TSF w/ UV with an additional info button that shows a static view of the historical performance of the forecasts of care demand and staff absences for the first one and a half weeks of the previous month. The descriptive text above translates to: ``\textit{Lookback forecast and actual care demand: This view compares forecasts from the previous month with the care demand that actually occurred. It is intended to help assess how reliable the forecasts are.}'' }
\end{subfigure}

\caption{Comparison of design variants TSF w/o UV, TSF w/ UV, and TSF w/ UV + PH.}
\label{fig:design_variants_tsf}

\end{figure}

\noindent\textbf{Time Series Forecasts}
The three design variants to answer RQ1 and RQ2 are visualized in~\Cref{fig:design_variants_tsf}. They all show different versions of the time series forecast interface (\Cref{sec:forecasts}), described in the following: 
\begin{description}
    \item[TSF w/o UV:] Time series forecast without uncertainty visualization
    \item[TSF w/ UV:] Time series forecast with uncertainty visualization
    \item[TSF w/ UV + PH:] Time series forecast with uncertainty visualization and historical performance visualization
\end{description}

\begin{figure}[t]
\centering

\begin{subfigure}{0.8\textwidth}
\centering
\includegraphics[width=\linewidth]{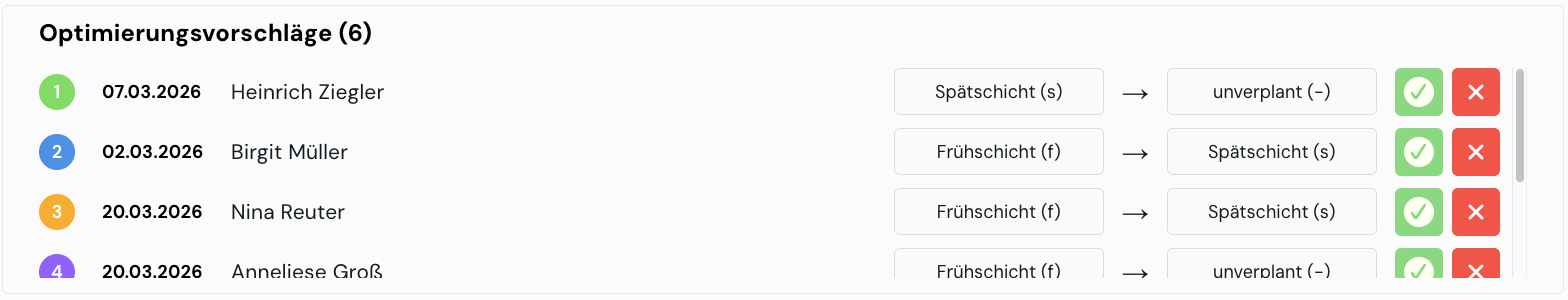}
\caption{AI Advice w/o Expl: AI advice without explanations.  For example, the system recommends assigning Heinrich Ziegler a day off instead of a on March 7, 2026.}
\end{subfigure}

\vspace{0.5cm}

\begin{subfigure}{0.8\textwidth}
\centering
\includegraphics[width=\linewidth]{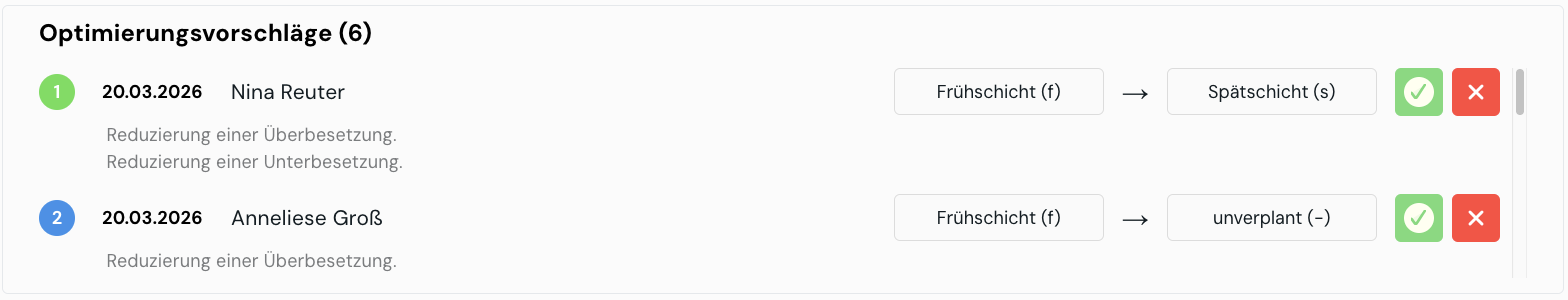}
\caption{AI Advice w/ Expl: AI advice with explanations. The grey text below every advice explains why a change is recommended. The first explanation the fictional nurse Nina Reuter translates to ``reduction of overstaffing, reduction of understaffing''.}
\end{subfigure}

\caption{Comparison of design variants AI Advice w/o Expl and AI Advice w/ Expl.}
\label{fig:design_variants:opti}
\end{figure}

\noindent\textbf{AI Advice}
To answer RQ3, two design variants of the AI advice interface (\Cref{sec:optimization}) were compared. Both interfaces showed six AI advice items, one with and one without explanations (see~\Cref{fig:design_variants:opti}), and participants were asked to accept or decline each piece of advice. In the following, we will refer to these designed variants as follows:
\begin{description}
    \item[AI Advice. w/o Expl.:] AI Advice without explanations
    \item[AI Advice w/ Expl.:] AI Advice with explanations
\end{description}

\begin{figure}
    \centering
    \includegraphics[width=1\linewidth]{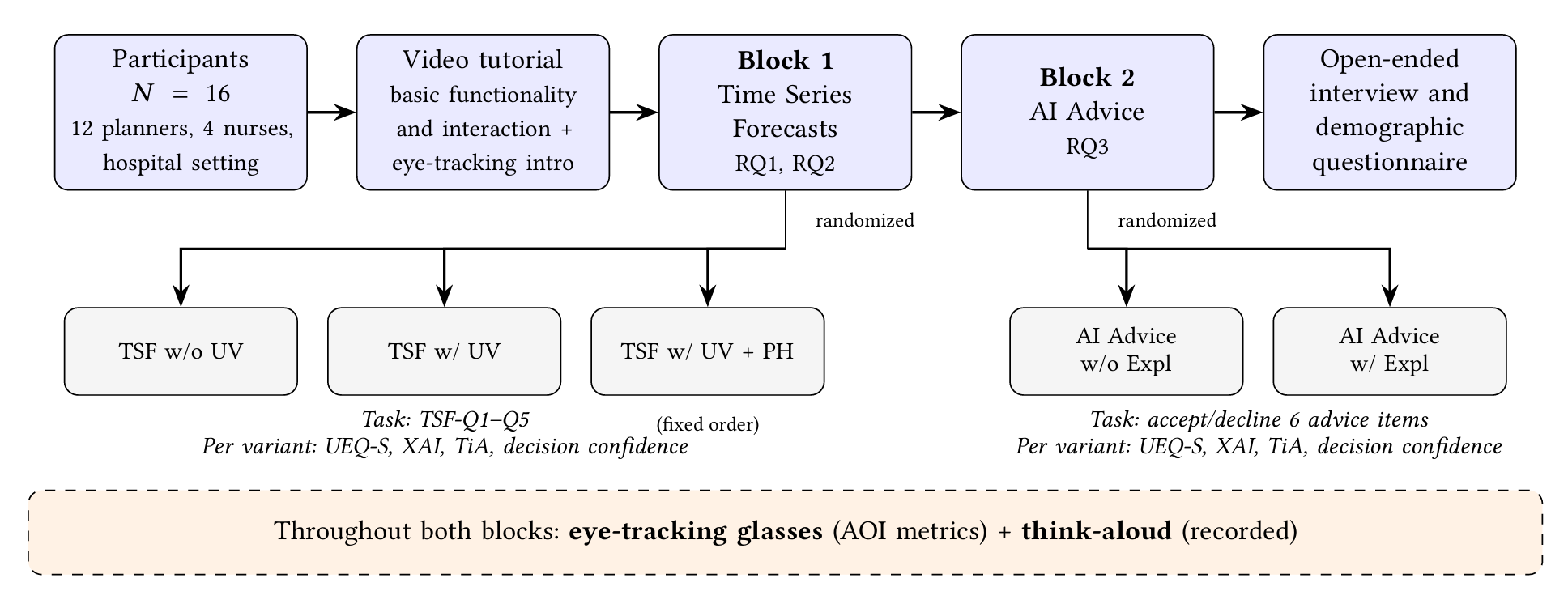}
    \caption{Within-subjects study procedure ($N=16$). Participants first watched a video tutorial with placeholder information for the design variants, then completed two sequential task blocks. \textbf{Block 1} (Time Series Forecasts; RQ1, RQ2) comprised three design variants, with TSF w/o UV and TSF w/ UV presented in randomized Latin-square order and TSF w/ UV + PH presented in fixed order. \textbf{Block 2} (AI Advice; RQ3) comprised two design variants presented in randomized order. After each variant, participants completed the UEQ-S, XAI scale, TiA questionnaire, and a decision-confidence rating. Eye-tracking and think-aloud data were collected throughout.}
    \label{fig:methodology}
\end{figure}

The study procedure is summarized in~\Cref{fig:methodology}. The uncertainty visualization task was conducted first, with the order of TSF w/o UV and TSF w/ UV randomized across participants. TSF w/ UV + PH was excluded from the randomization because this would have exposed participants to the uncertainty information before they encountered TSF w/ UV. The AI advice task followed, with the order of AI Advice w/o Expl and AI Advice w/ Expl also randomized. All randomized variants were balanced using a Latin-square-based approach.

The participants were asked the following questions during the interview to assess the understandability of the forecasts and the uncertainty visualization. All questions were restricted to the days between March 10 and 20 and the late shift:

\begin{description}
    \item[TSF-Q1] On which days was action required?
    \item[TSF-Q2] On which days is the care demand the lowest?
    \item[TSF-Q3] On which days is the uncertainty in the care demand forecast the lowest? (only for variants TSF w/ UV and TSF w/ UV + PH)
    \item[TSF-Q4] On which days are staff absences highest?
    \item[TSF-Q5] On which days is the uncertainty in the forecast of staff absences the highest? (only for variants TSF w/ UV and TSF w/ UV + PH)
\end{description}

After interacting with each design variant, participants completed a set of questionnaires. These included the UEQ-S~\cite{Schrepp_2017} to assess subjective user experience, a questionnaire assessing aspects of explainable AI~\cite{Hoffman_2023}, and the TiA questionnaire~\cite{McGrath_2025} to measure trust in automation. Thus, subjective evaluations were collected immediately after interaction with each respective variant.

Following the interaction tasks, participants took part in a short open-ended interview in which they were asked about positive and negative aspects of the overall experience, as well as to describe their typical daily workflow and how nurse scheduling is integrated into it.

At the end of the study, participants completed a demographic questionnaire assessing characteristics such as age, professional role, and prior experience with nurse scheduling and digital tools for nurse scheduling, with experience rated on a five-point scale.

\subsection{Study Setting}

\begin{figure}[t]
    \centering
    \includegraphics[width=0.7\linewidth, trim={0 0 0 3cm}, clip]{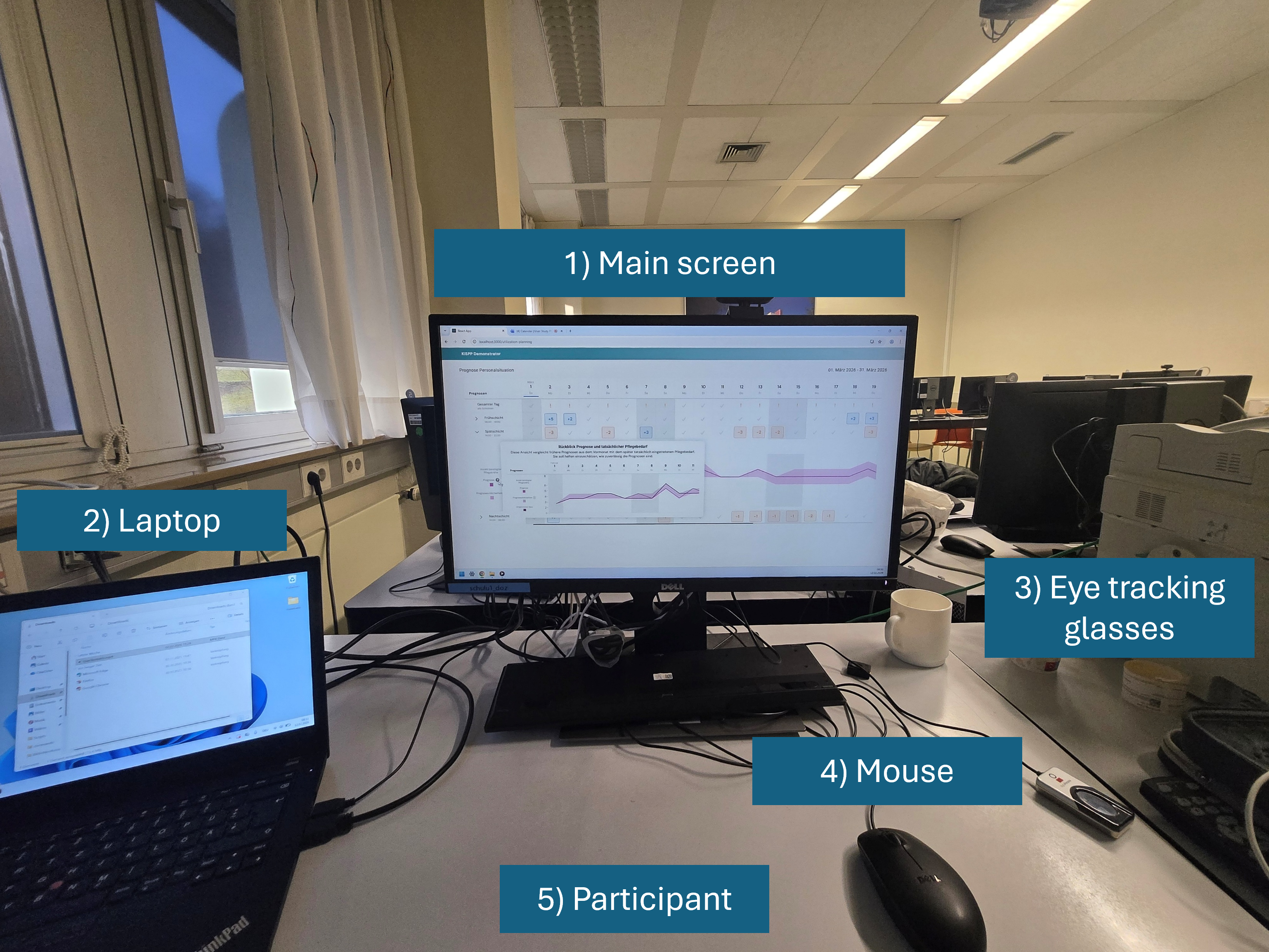}
    \caption{Study setting showing the main screen, the interviewer’s laptop, the eye tracking glasses, the mouse, and the participant’s position during the session.}
    \label{fig:StudySetting}
\end{figure}

Data collection was carried out in a prepared room in the hospital. 
Care was taken to ensure an environment with as few distractions as possible. The study was conducted via a main screen on which the tutorial, the answering of the questions, and the actual prototype of the personnel planning software were displayed. The participant was seated in front of the main screen and interacted with the system using a mouse. The interviewer used a laptop on the left side for technical purposes, but it was not part of the participant interface. For tasks not involving eye tracking, the glasses were placed on the table. Microsoft Teams was running in the background to record the thoughts verbalized during the interaction with the system and while answering the questionnaires. The study design is shown in the Figure \ref{fig:StudySetting}.

\subsection{Participants}

An on-site study was conducted at a hospital 
with 16 participants (age: M = 40.1 years, SD = 12.0) ranging in age from 28 to 64 years.. We made sure that we included a balanced number of male and female participants. However, we do not report the exact numbers for reasons of traceability. 12 participants held (assistant) ward manager roles (also called shift planners in the following) and therefore had frequent prior contact with nurse scheduling tools (self-rated experience with similar tools: M(SD) = 4.3(1.2); with nurse scheduling: M(SD) = 4.8(0.39)). Four participants were registered nurses and are therefore directly affected by the nurse scheduling process (self-rated experience with tools: M(SD) = 1.5(1.0); with shift scheduling: M(SD) = 2.5(1.3)). 


\subsection{Thematic Analysis}

We analyzed the interview data using inductive thematic analysis. After transcribing the interviews, we  extracted relevant excerpts from each participant. These excerpts served as the basis for an initial round of open coding. In collaborative sessions, our research team — four \ac{HCI}, \ac{AI}, and Requirements Engineering researchers — discussed and iteratively refined emerging codes. Through this process, we developed a preliminary codebook capturing recurring patterns across participants. We subsequently refined the codebook through repeated discussion. To support the development of higher-level themes, we  iteratively organized the excerpts into affinity-based clusters according to their conceptual similarities, following an affinity-diagramming-inspired process~\cite{Lucero_2022}.

To assess robustness, we reserved a subset of interviews for independent coding by all four researchers and calculated inter-rater reliability using Krippendorf's $\alpha$, applying a threshold of $>0.67$~\cite{marzi2024k}.  Initial scores fell below this threshold (0.436–0.62). After revising the codebook, scores remained insufficient (0.420–0.546). We therefore conducted a joint coding session to reconcile perspectives, which led to a simplified codebook and shared understanding. All remaining interviews were subsequently coded by at least two researchers according to the final codebook (see appendix). Excerpts assigned to the codes \textit{uncertainty communication}, \textit{system explanation of uncertainty visualization}, \textit{XAI AI advice}, and \textit{other XAI} were additionally annotated with sentiment (positive, neutral, or negative) toward these design elements.

\section{Results}

\begin{table}[t]
\caption{Performance comparison across tasks and design conditions. Mean Precision, Recall, and F1-scores (+- standard deviation) are reported across participants for each question and design variant within the first design block (time series forecasts). For questions involving three design variants, statistical differences were assessed using the Friedman test followed by post-hoc pairwise comparison where applicable. For questions involving two design variants, Wilcoxon signed-rank tests were used. Significant differences are indicated with an asterisk (*, p < 0.05). For each question and metric, the highest-performing design condition is highlighted in bold.}

\begin{tabular}{lllll}
\hline
Question (see~\Cref{task_questions})                 & Design & \multicolumn{1}{c}{Precision} & Recall                        & F1                           \\ \hline
\multirow{3}{*}{TSF-Q1}  & TSF w/o UV     & \textbf{0.99 +- 0.05}         & 0.94 +- 0.10                  & \textbf{0.96 +- 0.07}        \\
                      & TSF w/ UV     & 0.96 +- 0.08                  & 0.94 +- 0.17                  & 0.94 +- 0.13                 \\
                      & TSF w/ UV + PH     & 0.95 +- 0.12                  & \textbf{0.95 +- 0.15}         & 0.94 +- 0.13                 \\
                      \hline
TSF-Q2                   & TSF w/o UV     & 0.86 +- 0.34                  & 0.66 +- 0.35                  & 0.72 +- 0.34                 \\
                      & TSF w/ UV     & 0.88 +- 0.29                  & 0.60 +- 0.32                  & 0.70 +- 0.30                 \\
                      & TSF w/ UV + PH     & \textbf{0.91 +- 0.27}         & \textbf{0.73 +- 0.32}         & \textbf{0.79 +- 0.30}        \\
                      \hline
TSF-Q3                   & TSF w/ UV     & \textbf{0.41 +- 0.48}         & \textbf{0.41 +- 0.46}         & \textbf{0.38 +- 0.44}        \\
                      & TSF w/ UV + PH     & 0.28 +- 0.45                  & 0.28 +- 0.45                  & 0.28 +- 0.45                 \\
                      \hline
TSF-Q4                   & TSF w/o UV     & \textbf{0.75 +- 0.45}         & \textbf{0.75 +- 0.45}         & \textbf{0.75 +- 0.45}        \\
                      & TSF w/ UV     & 0.71 +- 0.45                  & \textbf{0.75 +- 0.45}         & 0.72 +- 0.45                 \\
                      & TSF w/ UV + PH     & 0.66 +- 0.46                  & \textbf{0.75 +- 0.45}         & 0.68 +- 0.45                 \\
                      \hline
TSF-Q5                   & TSF w/ UV     & 0.25 +- 0.32                  & 0.44 +- 0.51                  & 0.31 +- 0.37                 \\
                      & TSF w/ UV + PH     & \textbf{0.40 +- 0.32}         & \textbf{0.75* +- 0.45}         & \textbf{0.50 +- 0.35}
\end{tabular}
\label{tab:task_metrics_tsf}
\end{table}

\subsection{Objective Task Performance}

In the first design block on time series forecasting, we recorded the number of correct and incorrect responses to the comprehension questions (see~\Cref{task_questions}). As shown in~\Cref{tab:task_metrics_tsf}, Question TSF-Q1, which asked participants to identify the days requiring action, i.e., a change in the shift plan, was answered with consistently high precision, recall, and F1 scores across all design variants. In contrast, the questions pertaining to the forecast values (TSF-Q2 and TSF-Q4) yielded slightly lower scores. For TSF-Q2, design variant TSF w/ UV + PH achieved the highest average metric scores; however, no statistically significant differences were observed. For Question TSF-Q4, design variant TSF w/o UV yielded the highest precision and F1 score, again without statistical significance.

The questions asking participants to identify the days with the highest and lowest uncertainty (TSF-Q3 and TSF-Q5) received the lowest scores overall, suggesting that participants had difficulty correctly interpreting the uncertainty visualizations. The only statistically significant result was found for the recall of TSF-Q5, where participants in design variant TSF w/ UV + PH achieved a substantially higher average recall compared to design variant TSF w/ UV. Notably, for Question TSF-Q3, the pattern was reversed: performance was higher for design variant TSF w/ UV than for TSF w/ UV + PH, though this difference did not reach statistical significance.

\subsection{Eye Tracking Areas of Interest for AI Advice Designs}

To examine whether visual attention differed between design AI Advice w/o Expl and design AI Advice w/ Expl, two-sided paired Wilcoxon signed-rank tests were conducted separately for area of interest (AOI) 1 (AI Advices and, where applicable, their explanations), AOI 2 (rule violations), and AOI 3 (shift plans). For relative fixation duration, participants allocated a significantly larger proportion of their fixation time to AOI 1 in design AI Advice w/ Expl (Mdn = 23.70, IQR = 13.40) than to design AI Advice w/o Expl (Mdn = 18.80, IQR = 9.50), V = 114, p = .019, Holm-adjusted p = .037, $r_{rb}$ = .68. The relative fixation share for AOI 3 was significantly lower in design AI Advice w/ Expl (Mdn = 63.50, IQR = 15.90) than in design AI Advice w/o Expl (Mdn = 71.20, IQR = 11.90), V = 15, p = .007, Holm-adjusted p = .020, $r_{rb}$ = -.78. No significant difference emerged for AOI 2, V = 90, p = .266, Holm-adjusted p = .266, $r_{rb}$ = .32. 
Detailed descriptive eye tracking statistics are provided in Table \ref{tab:aoi-results}.

\begin{figure}[htbp]
    \centering
    \begin{subfigure}[t]{0.48\textwidth}
        \centering
        \includegraphics[
            width=\linewidth,
            trim=300mm 200mm 300mm 200mm,
            clip
        ]{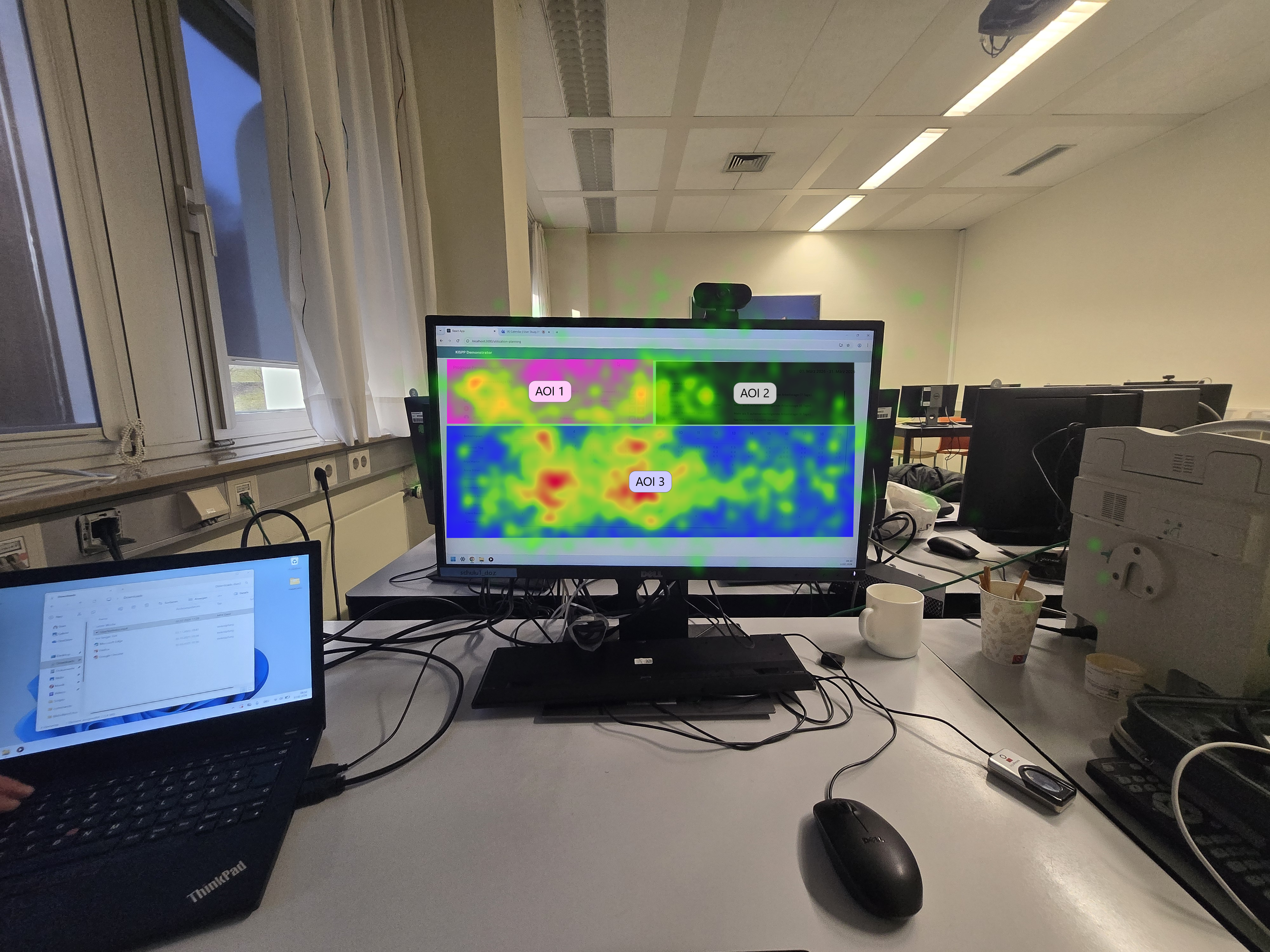}
        \caption{Design AI Advice w/o Expl without additional AI explanations}
        \label{fig:cooc_time_series}
    \end{subfigure}
    \hfill
    \begin{subfigure}[t]{0.48\textwidth}
        \centering
        \includegraphics[
            width=\linewidth,
            trim=300mm 200mm 300mm 200mm,
            clip
        ]{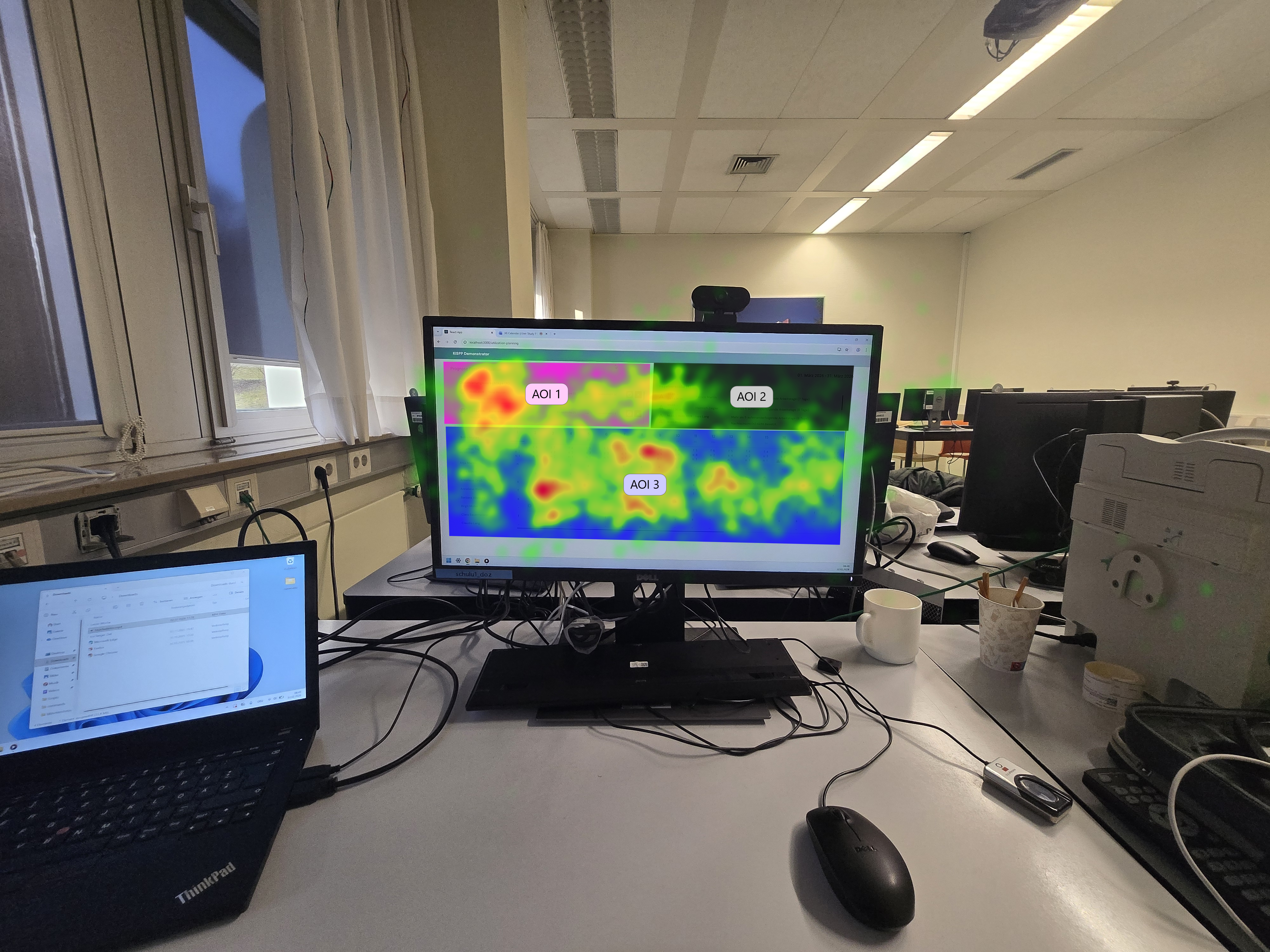}
        \caption{Design AI Advice w/ Expl with additional AI explanations}
        \label{fig:cooc_optimization}
    \end{subfigure}
    \caption{Comparison of relative fixation durations for AI Advice w/o Expl. (left) and AI Advice w/ Expl. (right). With additional explanations, visual attention is more concentrated in AOI 1 (upper left, AI advice and explanations) and less pronounced in AOI 3 (bottom, schedule).}
    \label{fig:heatmaps_eyetracking_relativ_fixations}
\end{figure}

This pattern is also clearly visible in Figure \ref{fig:heatmaps_eyetracking_relativ_fixations}, which depicts the relative fixation duration as a heatmap. Most notably, attention in design AI Advice w/ Expl increased substantially in the upper-left area of AOI 1, where the AI advice and explanations were presented. At the same time, AOI 3 (the shift plan) was viewed less intensely overall in design AI Advice w/ Expl than in design AI Advice w/o Expl, suggesting that the additional explanations directed attention more strongly toward the AI advice and reduced the need for manual visual verification in AOI 3.

These findings indicate that design AI Advice w/ Expl primarily altered the distribution of visual attention, shifting relative attention toward AOI 1 and away from AOI 3, rather than reliably changing the absolute time spent on the AOIs or the speed with which they were first fixated.

\begin{table}[htbp]
\centering
\caption{Descriptive eye tracking measures for the three areas of interest (AOIs) in design AI Advice w/o Expl and design AI Advice w/ Expl. Values are reported as means, with fixation duration (mean\_fix), time to first fixation (mean\_first\_fix), fixation share (mean\_share\_fix), number of fixations (mean\_n\_fix), and the number of participants whose first fixation occurred within each AOI (n\_fix). Each design group included \(n = 16\) participants.}
\label{tab:aoi-results}

\resizebox{\textwidth}{!}{%
\begin{tabular}{llrrrrrr}
\toprule
TOI & AOI &
mean\_fix (sec) &
sd\_fix (sec) &
mean\_first\_fix (sec) &
mean\_share\_fix (\%) &
mean\_n\_fix &
n\_first \\
\midrule

\multirow{3}{*}{Design AI Advice w/o Expl}
& AOI 1 & 28.31  & 20.50 & 19.39 & 18.73 & 156.00 & 6 \\
& AOI 2 & 10.51  & 7.61  & 13.29 & 8.13  & 57.63  & 3 \\
& AOI 3 & 123.71 & 81.98 & 3.48  & 73.15 & 637.81 & 7 \\

\midrule

\multirow{3}{*}{Design AI Advice w/ Expl}
& AOI 1 & 42.27  & 36.16 & 2.91 & 23.84 & 233.56 & 7 \\
& AOI 2 & 15.58  & 14.32 & 5.51 & 10.87 & 83.56  & 1 \\
& AOI 3 & 103.84 & 66.89 & 3.07 & 65.28 & 539.69 & 8 \\

\bottomrule
\end{tabular}%
}
\end{table}

\subsection{Decision Confidence in Relation to Task Performance and Acceptance of AI Advice}

To examine objective task performance, subjective decision confidence, and their relationship across designs, we analyzed both outcomes for TSF w/o UV, TSF w/ UV, and TSF w/ UV + PH. For each participant, objective task performance was calculated separately for each design as the mean F1 score across the three decision questions that were presented in all three TSF designs. Subjective decision confidence was assessed at the end of each task using a single-item 7-point rating scale, where higher values indicated greater confidence. All analyses accounted for the repeated-measures structure of the data.

\begin{figure}
    \centering
    \includegraphics[width=\linewidth, trim={0 0 0 0.8cm}, clip]{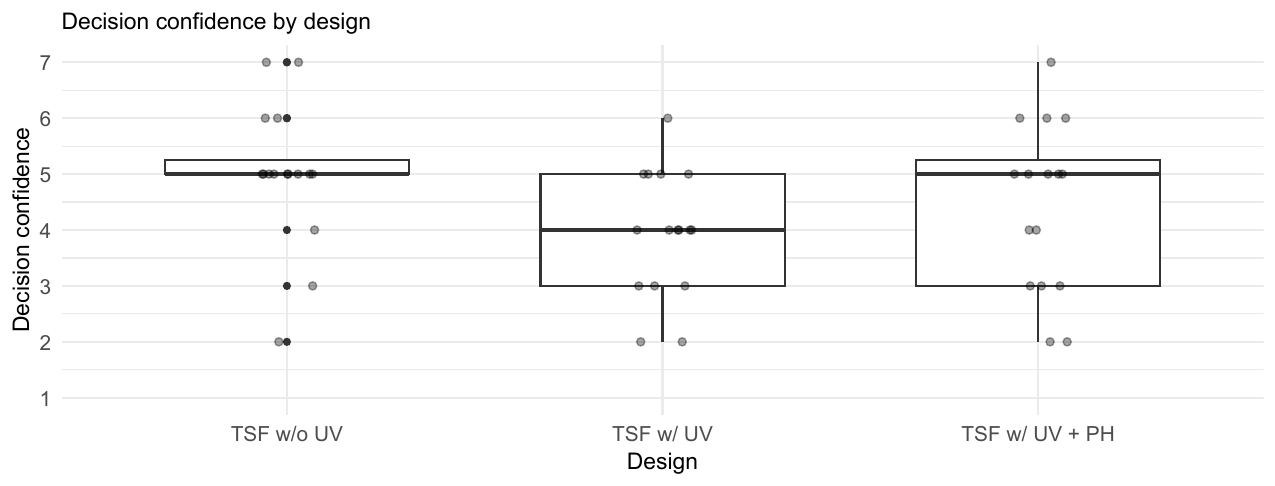}
    \caption{Decision confidence by design. Boxplots display decision confidence ratings for the three designs. Higher values indicate greater perceived confidence. Design TSF w/o UV included no uncertainty visualization, design TSF w/ UV included uncertainty visualization, and design TSF w/ UV + PH included uncertainty visualization plus history information.}
    \label{fig:Decision_Confidence}
\end{figure}

Across the three designs, decision confidence was highest for TSF w/o UV ($M = 5.00$, $SD = 1.26$), followed by TSF w/ UV + PH ($M = 4.44$, $SD = 1.50$), and TSF w/ UV ($M = 3.94$, $SD = 1.12$; see Figure~\ref{fig:Decision_Confidence}).

A linear mixed-effects model found no significant association between task performance and decision confidence, $b = -1.41$, $\mathit{SE} = 1.16$, $t(42.58) = -1.21$, $p = .232$, but revealed a significant main effect of design, $F(2, 21.36) = 5.68$, $p = .011$.

Holm-corrected pairwise comparisons of the estimated marginal means showed that decision confidence was significantly higher for TSF w/o UV than for TSF w/ UV, $M_{\mathrm{diff}} = 1.10$, $\mathit{SE} = 0.33$, $t(29.4) = 3.37$, $p_{\mathrm{Holm}} = .006$. No significant differences were found between TSF w/o UV and TSF w/ UV + PH, $M_{\mathrm{diff}} = 0.57$, $\mathit{SE} = 0.33$, $t(29.2) = 1.76$, $p_{\mathrm{Holm}} = .177$, or between TSF w/ UV and TSF w/ UV + PH, $M_{\mathrm{diff}} = -0.53$, $\mathit{SE} = 0.33$, $t(29.2) = -1.62$, $p_{\mathrm{Holm}} = .177$. Thus, the overall effect of design was primarily attributable to higher decision confidence in design TSF w/o UV compared to design TSF w/ UV.

For designs AI Advice w/o Expl and AI Advice w/ Expl, objective acceptance was operationalized as the number of accepted AI advice items, whereas subjective decision confidence was assessed using the same single-item 7-point scale. 
Descriptively, the two conditions were highly similar
(see Table~\ref{tab:desc_2ab_confidence}). Participants accepted an average of 4.44 advice items in design AI Advice w/o Expl ($SD = 1.75$), compared to 4.50 advice items in design AI Advice w/ Expl ($SD = 1.46$). Mean decision confidence was also comparable, with a mean of 5.69 ($SD = 0.79$) in design AI Advice w/o Expl and 5.75 ($SD = 1.06$) in design AI Advice w/ Expl.

Because the paired difference scores deviated from normality, differences between designs were examined using paired Wilcoxon signed-rank tests. Neither the number of accepted advice items, $V = 16.5$, $p = .729$, nor decision confidence, $V = 8.5$, $p = .890$, differed significantly between the two designs.

Descriptive correlations calculated separately for each design indicated a significant negative association between decision confidence and accepted AI advice items in design AI Advice w/o Expl, $\rho = -.669$, $p = .005$. Higher decision confidence was therefore associated with fewer accepted advice items. In design AI Advice w/ Expl, the association was also negative, but weaker and not statistically significant, $\rho = -.345$, $p = .190$. These separate correlations describe between-person associations within each design. They do not, by themselves, demonstrate that the strength of the relationship differed significantly between the two designs. To test this directly, a linear mixed-effects model was estimated with accepted items as the dependent variable and decision confidence, design, and their interaction as fixed effects. A random intercept for participants accounted for the repeated observations within individuals

\begin{figure}
    \centering
    \includegraphics[width=\linewidth, trim={0 0 0 0.6cm}, clip]{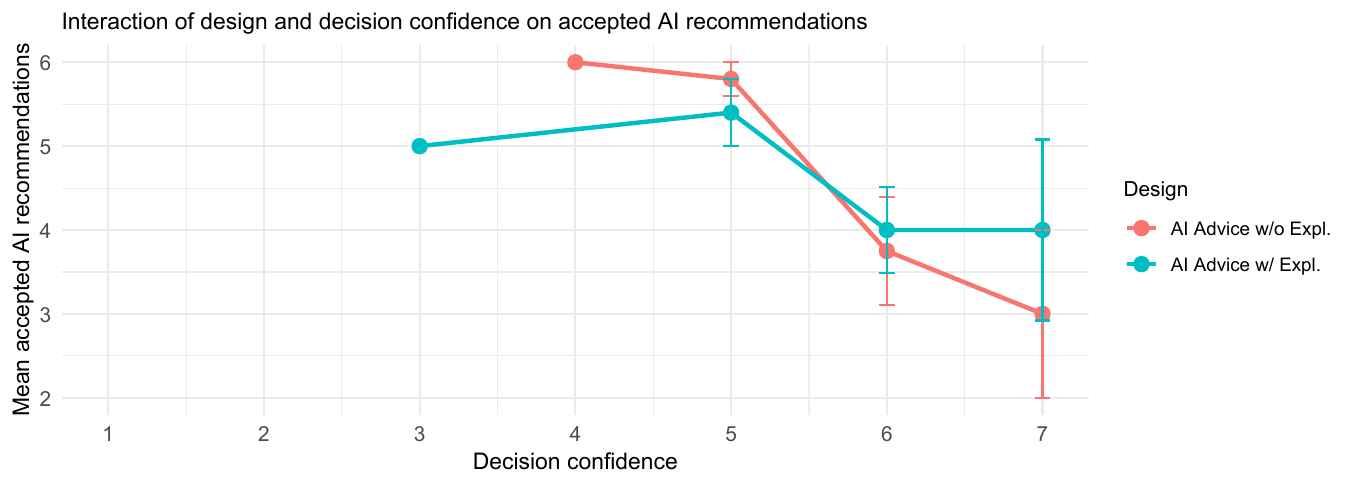}
    \caption{Interaction plot showing the relationship between decision confidence and the mean number of accepted AI advice items in designs AI Advice w/o Expl and AI Advice w/ Expl. Points represent mean values at each level of decision uncertainty, and error bars indicate standard errors. Design AI Advice w/o Expl presented AI advice without explanations, whereas design AI Advice w/ Expl presented AI advice with additional AI explanations.}
    \label{fig:Decision_Confidence_Interaction_2AB}
\end{figure}

The model revealed a significant effect of decision confidence, $F(1,31.36) = 5.68$, $p = .023$, with higher confidence associated with fewer accepted items. Neither the main effect of design, $F(1,14.92) = 0.14$, $p = .712$, nor the interaction between decision confidence and design, $F(1,17.00) = 2.18$, $p = .158$, were significant. Thus, although the negative association appeared descriptively stronger in design AI Advice w/o Expl, there was no statistically reliable evidence that the relationship differed between designs (see Figure~\ref{fig:Decision_Confidence_Interaction_2AB}).

\begin{table}[htbp]
\centering
\small
\caption{Descriptive statistics for accepted AI recommendations and decision confidence in designs AI Advice w/o Expl (n=16) and AI Advice w/ Expl. (n=16)}
\label{tab:desc_2ab_confidence}
\begin{tabular}{lrrrrrrrr}
\hline
Design 
& Accepted $M$
& Accepted $SD$
& Accepted $Md$
& Accepted $IQR$
& Confidence $M$
& Confidence $SD$ \\
\hline
AI Advice w/o Expl  & 4.44 & 1.75 & 5.00 & 3.00 & 5.69 & 0.79 \\
AI Advice w/ Expl  & 4.50 & 1.46 & 5.00 & 2.25 & 5.75 & 1.06 \\
\hline
\end{tabular}
\end{table}

\subsection{Questionnaire Results}

\paragraph{Comparison of UEQ-S, TiA, and XAI Across Designs.}
Each of the five design conditions comprised 16 participants. Separate comparisons were conducted for designs TSF w/o UV, TSF w/ UV, and TSF w/ UV + PH and for designs AI Advice w/o Expl and AI Advice w/ Expl. None of the comparisons revealed significant differences in the questionnaire results. Therefore, these results are not discussed in further detail. Descriptive statistics for the questionnaires are provided in the appendix.

\begin{figure}[t]
    \centering
    \includegraphics[width=\linewidth, trim={0 0 0 1,1cm}, clip]{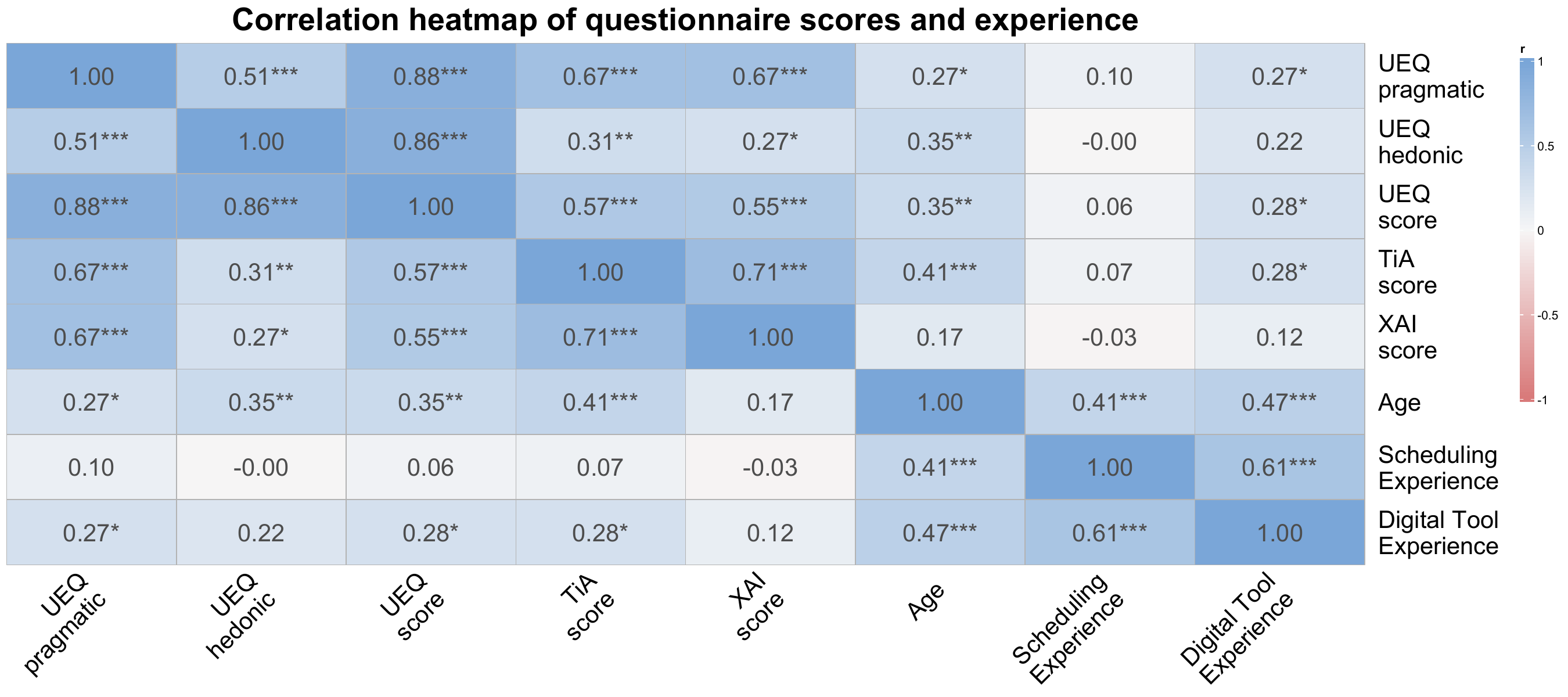}
    \caption{Correlation matrix of questionnaire measures assessing user experience (UEQ), trust in automation (TiA), and explainable artificial intelligence (XAI), alongside participants’ age and experience with scheduling and digital tools. Values represent Pearson’s correlation coefficients (r); blue colors indicate positive correlations, whereas red colors indicate negative correlations. Asterisks denote statistical significance (p < .05, *p < .01, **p < .001).}
    \label{fig:CorrelationMatrix}
\end{figure}

\paragraph{Correlations Between Questionnaire Measures and Participant Characteristics.}
The correlation analysis revealed several significant associations among user experience, trust in automation, perceived explainability, and participant characteristics (see \ref{fig:CorrelationMatrix}). As expected, the overall UEQ score was strongly correlated with both pragmatic quality (r = .88, p < .001) and hedonic quality (r = .86, p < .001), which were themselves moderately correlated ($r = .51$, $p < .001$). More positive overall user experience ratings were associated with higher trust in automation (r = .57, p < .001) and more favorable XAI ratings (r = .55, p < .001). Pragmatic quality showed particularly strong associations with both trust in automation and XAI ratings (r = .67, p < .001 for both). Hedonic quality was more weakly related to trust (r = .31, p < .01) and XAI ratings (r = .27, p < .05). Trust in automation and perceived explainability were strongly positively correlated (r = .71, p < .001).

Age was positively associated with pragmatic quality (r = .27, p < .05), hedonic quality (r = .35, p < .01), overall user experience (r = .35, p < .01), and trust in automation (r = .41, p < .001). However, age was not significantly correlated with XAI ratings (r = .17, p > .05). Age was also positively related to scheduling experience ($r = .41$, $p < .001$) and digital-tool experience ($r = .47$, $p < .001$), which were strongly correlated with each other ($r = .61$, $p < .001$). Digital-tool experience showed small positive associations with pragmatic quality (r = .27, p < .05), overall user experience (r = .28, p < .05), and trust in automation (r = .28, p < .05). In contrast, scheduling experience was not significantly associated with the questionnaire scores. 

Overall, the results indicate that more favorable perceptions of user experience and explainability were closely related to greater trust in automation. Higher age and greater digital-tool experience were associated with more positive user experience, pragmatic quality, and trust ratings.

\subsection{Findings from Interviews and Think-Aloud Protocols}
\label{sec:qualitative-findings}

\begin{figure}
    \centering
    \includegraphics[width=\linewidth]{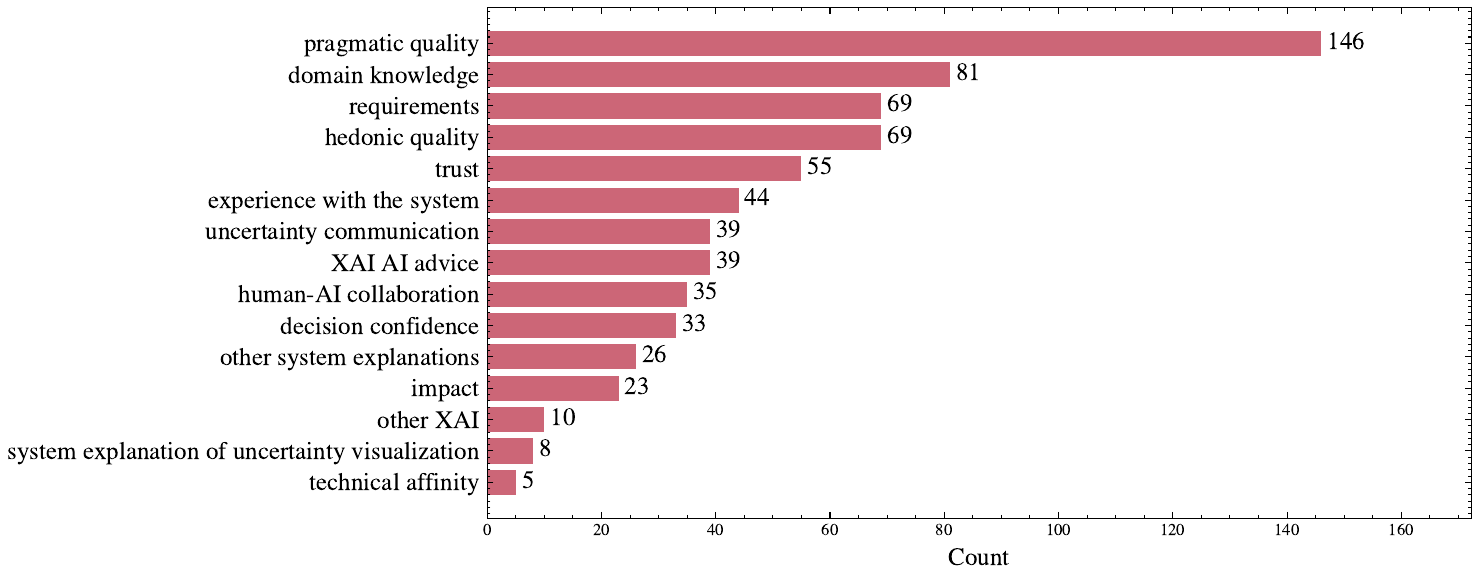}
    \caption{Total number of excerpts coded with a specific code.}
    \label{fig:code_frequency}
\end{figure}

\begin{figure}
    \centering
    \includegraphics[width=\linewidth]{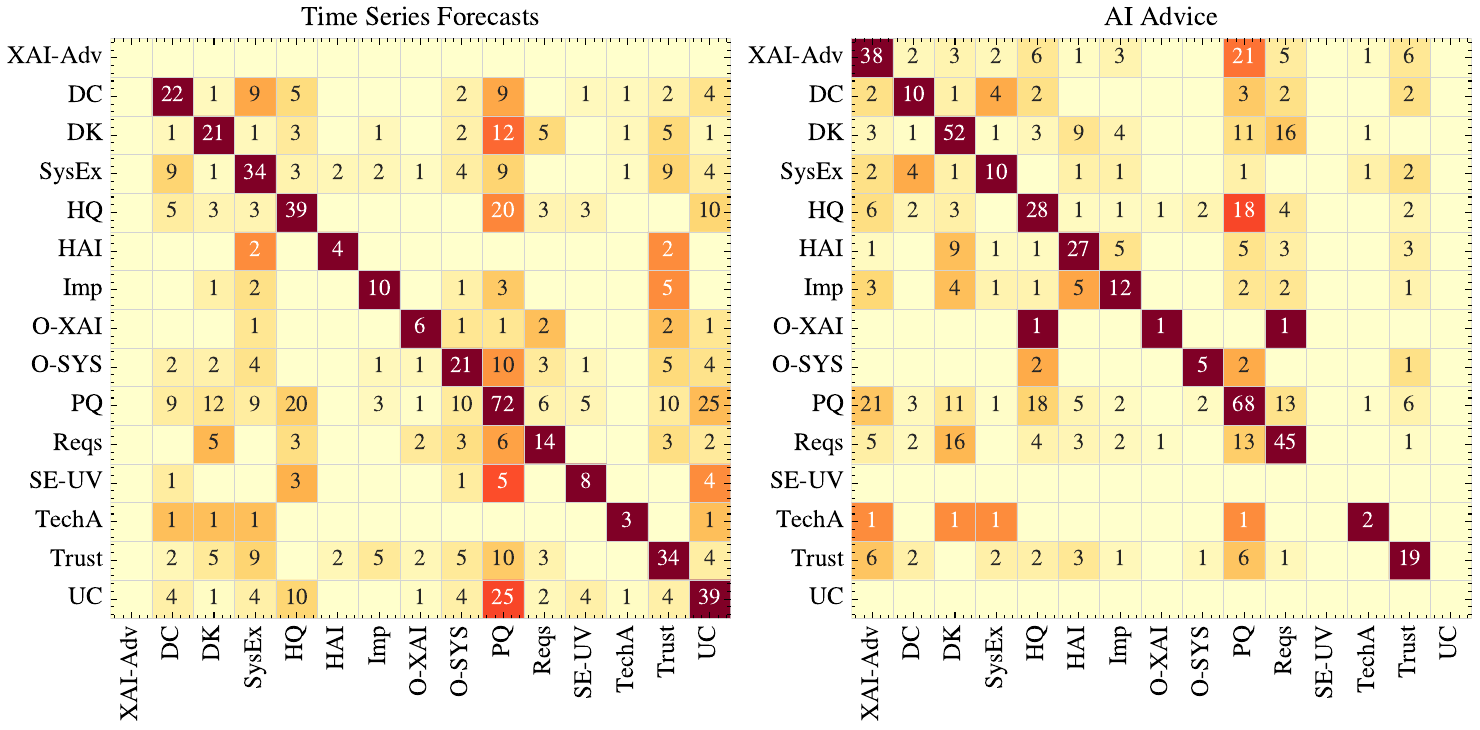}
    \caption{Code co-occurrence heatmaps per task block. Cell color reflects row-normalized values, indicating how frequently a given code co-occurred with another code relative to its total co-occurrences; all non-zero co-occurrences are indicated in rhe heatmap. The left panel shows co-occurrences for the \textit{Time Series Forecasts} block; the right panel for the \textit{AI Advice} block. Code abbreviations: \textit{XAI-Adv} = XAI AI Advice, \textit{DC} = decision confidence, \textit{DK} = domain knowledge, \textit{SysEx} = experience with the system, \textit{HQ} = hedonic quality, \textit{HAI} = human-AI collaboration, \textit{Imp} = impact, \textit{O-XAI} = other XAI, \textit{O-SYS} = other system explanations, \textit{PQ} = pragmatic quality, \textit{Reqs} = requirements, \textit{SE-UV} = system explanation of uncertainty visualization, \textit{TechA} = technical affinity, \textit{Trust} = trust, \textit{UC} = uncertainty communication.}

    \label{fig:Cooc_per_zone_grid}
\end{figure}

The qualitative analysis yielded a codebook of fifteen codes organized across two task blocks
(see Figure~\ref{fig:code_frequency} for code frequencies;
Figure~\ref{fig:Cooc_per_zone_grid} for co-occurrence patterns).
The most frequently applied codes across both blocks were \textit{pragmatic quality} (PQ),
\textit{domain knowledge} (DK), \textit{requirements} (Reqs), and XAI and uncertainty communication-related codes.
The following subsections report findings structured around the three research questions:
Uncertainty comprehension and interpretation (RQ1), the influence of uncertainty visualization on decision confidence (RQ2), and the influence of
AI advice explanations on acceptance and decision confidence (RQ3), followed by cross-cutting findings on \textit{trust}, \textit{decision confidence}, and \textit{human-AI collaboration}.

\subsubsection{Uncertainty Comprehension and Interpretation (RQ1)}
\label{sec:findings-rq1}

Uncertainty communication was among the most discussed topics in the time series forecast task block and elicited predominantly negative sentiment.
Of the 44 citations coded as either \textit{uncertainty communication} (UC) or \textit{system explanation of uncertainty visualization} (SE-UV) (see~\Cref{fig:design_variants_tsf} (b)), only five carried positive sentiment, while 23 were negative and 16 neutral. Nearly all negatively coded citations co-occurred with \textit{pragmatic quality} (PQ: $n=25$), indicating that participants repeatedly questioned the practical utility of the uncertainty visualization.

Participants struggled particularly when uncertainty bands were displayed alongside the forecast line, with some conflating the prediction interval with the point forecast itself.
P4 summarized a widely shared reaction:
\textit{``Far too much information. And those lines – they make it far too confusing.''} and \textit{``But what does this line actually mean? [\ldots] What is the difference between this forecast uncertainty and the forecast?''}). A simpler, single-line visualization was preferred by P9:
\textit{``I’d just like a simple line that fluctuates up and down [\ldots] – just to show trends .''} Additionally, when asked to identify days of highest or lowest uncertainty, participants frequently looked at the peak or trough of the prediction interval rather than its width
(P5: \textit{``I would have said that [this day has the highest uncertainty], because the forecast uncertainty indicates a peak there.''}).
These patterns persisted even when an explanatory note on how to read the visualization was provided, suggesting that the explanatory information of how to interpret the uncertainty visualization was insufficient.

\subsubsection{Uncertainty Visualization: Actionability and Decision Confidence
(RQ2)}
\label{sec:findings-rq2}

Beyond interpretation difficulties, uncertainty communication also shaped participants' willingness to engage with the forecast view and their confidence in acting on it.

\paragraph{Actionability Gap.}
Even when the uncertainty visualization was broadly understood, participants questioned what to do with the information. P7 raised: \textit{``And the other question is, what do I do with the uncertainty??''}
The perceived utility of the uncertainty visualization was experience-dependent: P10 noted that it offered limited value to experienced planners
(\textit{``Let’s be honest, as a shift planner, it’s not much use to me if the program says that it's uncertain''}),
while P2 suggested it would be especially useful during onboarding (\textit{``Especially at the start, one would definitely make use of this view frequently.''}).
Where actionability was articulated, it connected to \textit{domain knowledge} (DK): P10 described how \textit{``high uncertainty on absence days would prompt planning a personnel buffer''}.

\paragraph{Demand for XAI and Explanations of Uncertainty.}
Several participants ($n=9$) expressed a desire for explanatory information in the time series forecast designs. P2 articulated this clearly: \textit{``So why that particular date,  5 January? It would help to have an explanation as to why the system chose that day.''} This finding suggests that the demand for XAI extends beyond the AI advice in the shift plan to the forecast functionality. P1 also expressed a desire to understand \textit{why} uncertainty varies across days
(P1: \textit{``It makes me wonder why, on some days, it is possible to calculate exactly when an absence will occur and on others it is not?''}).

\subsubsection{Explainability of AI Advice: Acceptance and Decision Confidence (RQ3)}
\label{sec:findings-rq3}

\textit{XAI AI Advice} (XAI-Adv) was among the most frequently
coded theme in the optimization task block (XAI-Adv: $n=38$). The sentiment of excerpts labeled with XAI-Adv and \textit{other XAI} (O-XAI) is predominantly positive with 32 positive, four negative, and thirteen neutral statements. XAI-Adv co-occurred most strongly with \textit{pragmatic quality} (PQ: $n=21$), \textit{trust} ($n=16$), and \textit{domain knowledge} (DK: $n=13$), indicating that explanations were evaluated at the intersection of usability, trust, and professional context.

\paragraph{Positive Reception and Increased Willingness to Use.}
Most participants ($n=11$) who actively engaged with the explanations found them meaningful and useful, and several ($n=5$) reported increased willingness to act on the
AI advice. Participants valued knowing the rationale behind an advice item, particularly when it aligned with or extended their domain knowledge. P1 noted that explanations would be especially valuable for less experienced shift planners.

\paragraph{Cognitive Overload and Salience.}
Only three participants raised negative points regarding the explanations, all related to information overload (e.g., P16: "I think there’s far too much text again").
A few participants did not notice the explanations at all during the interaction (P4: \textit{``No''} when asked whether they had seen the explanations), pointing to a salience problem in the current design. This tension between transparency and cognitive economy was illustrated by P7's preference for the version without explanations: \textit{``Less is more.''}

\subsubsection{Trust and Decision Confidence}
\label{sec:findings-trust}

\textit{Trust} merged as a prominent cross-cutting theme, co-occurring frequently with both \textit{XAI AI Advice} (XAI-Adv) and \textit{uncertainty communication} (UC).

\paragraph{Stakes-Driven Distrust.}
A recurring source of distrust was the perceived risk of acting on incorrect AI outputs in the high-stakes professional context ($n=3$).
One participant articulated how blind reliance on the system could propagate planning errors that ultimately affect patient safety.
This pattern illustrates how professional responsibility amplifies distrust: the
perceived cost of an AI error is not merely operational inconvenience but potential harm to patients and nursing staff. P8, in contrast, expressed unconditional trust:
\textit{``I’m not skeptical at all. Harmful or adverse consequences? I don’t think so. I trust the system.''}

\paragraph{Explainability as a Trust Enabler.}
Several participants reported increased trust when explanations accompanied AI advice. P5 confirmed directly: \textit{Moderator: ``Did the explanations help to improve trust in the AI as well??'' P5:``Yes.''} P10 echoed this: \textit{``I'm skeptical about the system. Well, I would definitely need the instructions [explanations].''} Conversely, P4 described early mistrust stemming from a lack of explanation:
\textit{``Well, I think at first you are just a bit skeptical when you are are not familiar with it [\ldots], I mean, why does the forecast say there will be increased staff absences on 16 March?''} 

\paragraph{Learning Curve and Growing Confidence.}
Repeated interaction with the system reduced initial confusion and increased decision confidence. P4 noted:
\textit{``But the good thing is, by the third time, I was already feeling a little more confident.''} P8 similarly reported that initial confusion about AI advice resolved with use: \textit{``At first, I didn't understand it—not even this AI advice. Now I get how it's supposed to work.''}

\subsubsection{Human-AI Collaboration}
\label{sec:findings-hai}

Across both task blocks, participants consistently framed the tool's role as assistive rather than autonomous. In the co-occurrence analysis, \textit{human-AI collaboration} (HAI) appeared frequently alongside \textit{domain knowledge} (DK) (Figure~\ref{fig:Cooc_per_zone_grid}), indicating that views on collaboration were tightly coupled to professional expertise and awareness of social dimensions that AI cannot fully capture.

\paragraph{Human Oversight as Professional Necessity.}
Participants repeatedly described AI advice as a starting point rather than a means to fully automate scheduling (e.g., P5: \textit{``Well, I really think that it's better to build a basic structure and then make adjustments [with the AI] than to have a finished plan [created by the AI].''}). Oversight was not framed as distrust per se, but as a professional necessity given the social complexity of nurse scheduling: P5: \textit{``Several colleagues have a lot of wishes [\ldots], so I think it's important for a human to take another look at it''}, and P4: \textit{``Of course, a human always has to double-check. Especially in nursing.''}

\paragraph{Efficiency Benefits and Workflow Integration.}
Participants acknowledged the efficiency benefits of AI assistance for time-consuming tasks, e.g. P4: \textit{``I think the AI takes a few seconds, and you'll need—well, if you really sit down and do it—a whole day, a day and a half.''} Several participants ($n=8$) explicitly endorsed an AI-supported workflow, provided that individual preferences and fairness constraints could be incorporated into the optimization (e.g., P6: \textit{``If, for example, the AI could also include personal data as rule violations [\ldots], I would actually think that would be really cool.''}). Overall, participants viewed AI as an extension under human supervision rather than a full delegation of tasks.

\section{Discussion}
\label{sec:discussion}

\subsection{How well is the uncertainty information understood, and to what extent do brief explanations of how to interpret the uncertainty information improve understanding and interpretation? (RQ1)}
\label{sec:discussion-rq1}

\begin{tcolorbox}[colback=gray!15, colframe=gray!15, boxrule=0pt, arc=4pt]
\textbf{Findings:}
Uncertainty communicated as prediction intervals was poorly understood. Participants systematically confused the uncertainty band width with the peak or trough. Brief in-interface explanations were insufficient to remedy misinterpretation.
\end{tcolorbox}

Across all data sources, the findings converge on a clear picture: uncertainty visualized as prediction intervals is poorly understood by professional nurses and shift planners, and brief explanations of the visualization are insufficient to remedy this.

Objective task performance was lowest precisely for the uncertainty-specific questions (TSF-Q3 and TSF-Q5), with mean F1-scores well below those for basic forecast comprehension tasks. The think-aloud data reveal the dominant misinterpretation pattern: participants identified days of highest or lowest uncertainty by the peak or trough of the band rather than by its width. These errors persisted even after participants had read the explanatory note on how to interpret the visualization. This finding is inconsistent with prior work where users understood prediction interval visualizations without prior explanation~\cite{vis_uncertainty, joslyn2013benefits}.

Participants also struggled to act on uncertainty information. Even when they understood the visualization, they questioned both the source of that variability and its practical implications for shift planning. Positive task performance on Question TSF-Q1, TSF-Q2, and TSF-Q4 (identifying days requiring action and identifying days with low care demand or high staff absences) confirms that participants could reliably extract actionable signals from the forecasts; the difficulty was specific to the uncertainty layer, not to the forecast as a whole.

\subsection{How does uncertainty visualization influence users’ confidence in decision-making? (RQ2)}
\label{sec:discussion-rq2}

\begin{tcolorbox}[colback=gray!15, colframe=gray!15, boxrule=0pt, arc=4pt]
\textbf{Findings:}
Uncertainty visualization significantly reduced decision confidence ($F(2, 21.36) = 5.68,\ p = .011$). Participants reported higher confidence without uncertainty visualization than with it.
\end{tcolorbox}

Uncertainty communication had a measurable negative influence on decision confidence ($F(2, 21.36) = 5.68,\ p = .011$). Participants reported significantly higher confidence in design TSF w/o UV (forecast without uncertainty) than in design TSF w/ UV (forecast with uncertainty). This finding stands in contrast to prior work reporting beneficial effects of prediction interval visualizations on decision confidence and advice utilization~\cite{joslyn2013benefits,vis_uncertainty,leffrang2021should}. A plausible explanation lies in the specialized user group and domain context, as prior work confirms that such effects are highly domain and visualization dependent~\cite{hulman2019}. 

Design TSF w/ UV + PH, which added historical forecast performance, did not significantly improve decision confidence or trust over either design TSF w/o UV or design TSF w/ UV. UEQ-S scores did not differ significantly across the three TSF designs, suggesting that overall user experience was not critically harmed by communicating uncertainty; the frustration was specific to the uncertainty layer rather than the tool as a whole.

Notably, objective task performance was not significantly related to decision confidence ($b = -1.41$, $p = .232$), meaning differences in confidence between designs cannot be explained by differences in task performance, and confidence was not a reliable indicator of decision quality.

\subsection{How do explanations of AI recommendations influence users’ acceptance and their confidence in decision-making? (RQ3)}
\label{sec:discussion-rq3}

\begin{tcolorbox}[colback=gray!15, colframe=gray!15, boxrule=0pt, arc=4pt]
\textbf{Findings:}
Explanations were subjectively valued and shifted visual attention toward the explanation area, yet did not significantly change acceptance rates or decision confidence. 
\end{tcolorbox}

The findings reveal an important divergence between subjective attitudes and objective behavior. On the subjective side, qualitative data indicate that explanations were broadly welcomed. Several participants ($n=4)$ explicitly connected explanations to greater willingness to act on AI advice and higher trust.
Trust and perceived explainability were strongly correlated ($r = .71$, $p < .001$), and qualitative data confirm that explanations were a mechanism by which initial mistrust was reduced (see~\Cref{sec:findings-trust}). Eye-tracking data provide corroborating behavioral evidence: in design AI Advice w/ Expl, participants allocated significantly more fixation time to the explanation area (AOI1; $r_{rb} = .68$, Holm-adjusted $p = .037$) and correspondingly less to the shift plan area (AOI3; $r_{rb} = -.78$, Holm-adjusted $p = .020$). 

On the objective side, however, design AI Advice w/ Expl did not produce significantly higher questionnaire scores (UEQ-S, XAI scale, TiA), nor did it change the number of accepted advice items or decision confidence.  A negative association between decision confidence and acceptance rate was present in both designs, and the interaction effect was not significant, providing no reliable evidence that explanations changed this relationship.

The eye tracking pattern suggests that explanations \textit{substituted for} rather than supplemented independent verification: attention shifted toward the explanation and away from the shift plan on which the decision was made. Together with unchanged acceptance rates, this indicates a shallow, heuristic mode of engagement in which the explanation served as a reassurance cue rather than as evidence to be evaluated. Such engagement is consistent with the effect of ``placebic explanations''~\cite{eibandImpactPlacebicExplanations2019}, derived from ``placebic information''~\cite{langer1978mindlessness}, where the mere presence of a justification raises perceived trust independently of its content, and it is precisely the mode that cognitive forcing designs are intended to disrupt~\cite{bucincaTrustThinkCognitive2021}.

The salience problem identified in the qualitative data is a likely contributing factor: Three participants missed the explanations entirely, and others ($n=2$) found them too text-heavy to engage with fully. This suggests that the current implementation failed to make explanations sufficiently salient and concise to reliably influence decision behavior.


\subsection{Reflection of Practical and Societal Impact}
\label{sec:practical_societal_impact}

\paragraph{Trust calibration in a safety-critical domain.}
In the cross-sectional correlational analysis, perceived explainability was the strongest predictor of trust in automation ($r = .71$, $p < .001$), stronger than pragmatic quality ($r = .67$), hedonic quality ($r = .31$), or experience-related variables. Pragmatic quality, in turn, was strongly associated with both trust and XAI ratings ($r = .67$ for both), suggesting that trust is built primarily through the functional experience of the tool. The qualitative data add a longitudinal dimension to this picture: participants described trust and decision confidence as developing over time through repeated interaction, consistent with process models of human-automation trust that distinguish between initial calibration and experience-based recalibration~\cite[e.g.,][]{hoff2015trust}. The finding that digital-tool experience — but not scheduling experience — was positively associated with trust ratings suggests that familiarity with similar automated systems, rather than domain expertise, shapes the starting point of this calibration process. The high-stakes nature of nurse scheduling amplifies both the importance and the difficulty of appropriate trust calibration. Four participants articulated heightened awareness of the consequences of AI errors for staff and patients, a healthy response in this domain. However, growing familiarity with the tool may also carry risks: some participants acknowledged that repeated interaction could lead to increasing trust, potentially leading to overtrust and thus in overreliance. One participant even mentioned to place unconditional trust in the system (see~\Cref{sec:findings-trust}). 

\paragraph{Fairness in AI-assisted scheduling.}
Deploying AI-assisted scheduling in hospitals carries implications that extend beyond usability and decision support effectiveness. In a profession already characterized by high workload and power imbalances~\cite{griffiths2020nursing,RizanyNurseSatisfactionScheduling,stimpfel2025scheduling}, nurses risk being systematically disadvantaged by optimization systems that treat all inputs as equivalent. Our qualitative data — consistent with prior work~\cite{renggli_integrating_2025,userstudy} — reveal that participants want scheduling systems to incorporate and fairly weigh individual staff wishes. Yet treating all preferences equally does not necessarily yield a fair schedule~\cite{fuchs2025roadmap}: fairness may additionally require equitable distributions of preference satisfaction, workload, and accumulated overtime across staff. Future deployments should therefore allow such contextual inputs to be captured, made accessible, and subjected to transparent human review, while maintaining data privacy.

A related risk concerns how explanations shape fairness perceptions. Prior research shows that XAI can increase perceived fairness even when the underlying system is unfair, and that explanations can foster trust independently of their actual content~\cite{deckCriticalSurveyFairness2024}. In our context, explanations accompanying scheduling advice could lead planners and nurses to perceive a schedule as more equitable than it actually is, a form of miscalibrated fairness perception that may mask rather than surface problematic optimization behavior. As Deck et al.~\cite{deckCriticalSurveyFairness2024} argue, the goal of XAI in high-stakes settings should be to support \textit{calibrated} fairness perceptions rather than merely positive ones. Designed inclusively, such a system could actively reduce, rather than strengthen, the inequities that already arise in manual scheduling.

\paragraph{Risks of institutional misuse.}
AI-optimized scheduling outputs may also be appropriated at an institutional level in unintended ways. P9 raised the concern that HR could use optimized schedules to justify staff reductions — and while they assumed this would not occur, the risk is plausible. If AI outputs are taken to demonstrate that demand can be met with fewer nurses, decision-makers may cut positions, compromising patient safety and staff wellbeing~\cite{stimpfel2025scheduling}. These outcomes are in direct contradiction to the intended purpose of the system.

\paragraph{Human oversight and participatory design.}
Participants consistently framed human-AI collaboration as augmentation under human oversight, not as a lack of trust, but as a professional and ethical necessity given the social complexity of shift scheduling (see~\Cref{sec:findings-hai}). This is consistent with prior qualitative evidence that nurses and other healthcare professionals view AI decision support as an aid to, rather than a substitute for, clinical judgment~\cite{nurses_ml,sivaraman2023ignore,renggli2025integrating}. Scheduling algorithms should support, not replace, the professional judgment and negotiation between staff that this task inherently requires. We therefore view participatory design — engaging not only direct users but also affected parties such as nurses — as a prerequisite for responsible deployment.

\subsection{Design Implications}
Based on the results and discussion from the studied nurse scheduling use case, we derived the following design implications for AI-based decision support systems, particularly in planning and scheduling contexts:

\paragraph{\textbf{Design Implication~1: Replace static prediction intervals with interaction-first uncertainty representations.}}
Given the persistent misinterpretation of prediction interval bands and the negatively considered user experience with respect to the uncertainty visualization (see~\Cref{sec:discussion-rq1}), designers should consider alternative uncertainty representations that are more compatible with non-expert chart literacy, such as color-encoded risk categories, textual warnings of high forecast uncertainty above a certain threshold, or progressive disclosure designs in which uncertainty detail is revealed only on demand to reduce cognitive overload.

\paragraph{\textbf{Design Implication~2: Use examples to clarify uncertainty communication.}}
Given that the misinterpretation errors persisted even when participants had read the explanatory note on how to interpret the visualization (refer to~\Cref{sec:discussion-rq1}), interfaces with uncertainty communication should use clear examples to demonstrate how the information should be interpreted.

\paragraph{\textbf{Design Implication~3: Embed actionable guidance directly in the uncertainty visualization.}}
Because participants understood uncertainty information better when it was linked to concrete planning actions (e.g., scheduling a personnel buffer on high-uncertainty
days) (see~\Cref{sec:discussion-rq1}, and~\Cref{sec:discussion-rq2}), the visualization should surface explicit recommendations rather than leaving the translation to the user. This could take the form of contextual annotations that suggest specific planning responses when uncertainty exceeds a defined threshold. This reframing connects the uncertainty display to the planner's existing workflow and domain knowledge, reducing the cognitive gap between perceiving and acting on uncertainty.

\paragraph{\textbf{Design Implication~4: Prioritize salience and conciseness over completeness in AI advice explanations.}}
The finding that some participants missed the explanations entirely, while others found them cognitively overwhelming (\Cref{sec:discussion-rq3}), points to a fundamental design challenge: the current textual format placed the full cognitive burden on the user.
Explanations should be redesigned for progressive disclosure: a brief, prominent rationale by default, with fuller detail available on demand, preferably directly in the shift plan where the decision is made. Visual cues such as color coding, iconography, or inline highlighting could improve salience without increasing text density.

\paragraph{\textbf{Design Implication~5: Extend explanatory information to forecast visualizations.}}
Participants spontaneously requested explanations not only for the AI advice but also for the time series forecasts, asking why specific days were flagged and why uncertainty varied across the planning horizon.
This demand is currently unmet in the design. Future iterations should provide on-demand explanations for forecast outputs, linking high uncertainty or high forecast values to identifiable drivers such as historical patterns, known staffing constraints, or seasonal effects.

\paragraph{\textbf{Design Implication~6: Design for calibrated trust, not maximum trust.}}
Given that the perceived costs of AI errors in nurse scheduling can extend to harm to patients, and given that our findings suggest that some users tend to have an unconditional trust in the AI’s advice (e.g.,~\Cref{sec:practical_societal_impact}), the system should actively support appropriate skepticism alongside positive engagement. Therefore, the goal should not be to maximize trust, but to support \textit{calibrated} trust to achieve appropriate reliance~\cite{schemmerAppropriateRelianceAI2023} — actively encouraging ongoing scrutiny rather than passive acceptance. This could include transparent communication of the AI’s known limitations and historical accuracy, explicit options for human review and override, and verification mechanisms that enable planners to understand the consequences of accepted advice.

\paragraph{\textbf{Design Implication~7: Design AI scheduling tools as collaborative tools with explicit options for manual override.}}
The user interface should make it clear at every stage that the AI provides suggestions rather than making decisions, and that final decision authority remains with the human user. The options available to the user for modifying and reordering advice items, and for explaining why they have accepted or rejected them, should be clearly visible and easy to use. This helps ensure that the AI remains a decision-support tool while preserving human control and allowing users to incorporate their workflow requirements and domain knowledge into the final scheduling decision.

\paragraph{\textbf{Design Implication~8: Incorporate staff preferences and fairness constraints as optimization inputs.}}
To move toward human–AI collaboration that users perceive as useful and fair, future systems should provide structured mechanisms for planners and nurses to specify individual staff preferences, workload equity constraints, and other social considerations as optimization inputs rather than post-hoc corrections, thereby incorporating users’ domain knowledge and situated needs directly into the planning process.

\subsection{Study Limitations}

Some limitations should be considered when interpreting the findings. First, the sample consisted of only 16 participants, limiting statistical power and generalizability. However, the study involved a specialized and difficult-to-recruit target group and was conducted in a real-world hospital setting with actual hospital staff. It therefore prioritized context-specific expert perspectives over a larger sample. Nevertheless, the findings should be interpreted cautiously and validated in larger, multi-site studies.

Second, uncertainty was visualized exclusively through prediction intervals. Consequently, the findings may not generalize to other uncertainty-visualization approaches, such as frequency-based representations or hypothetical outcome plots.

Although participants were recruited from a single hospital, we deliberately included individuals with varying levels of experience and aimed for a balanced distribution across genders and age groups. Nevertheless, differences in organizational practices and workflows may limit transferability to other clinical settings.

Finally, the mock data did not always correspond to the actual ward sizes of the participants or to specific organizational rules, such as the maximum number of consecutive working days. These inconsistencies may have caused confusion and reduced the ecological validity of the study. Care was taken, however, to ensure that the simulated data resembled realistic data. The use of real data from one of the hospital’s wards might have led to increased cognitive load due to the large number of staff members.

\section{Conclusion}

This paper presented a mixed-methods user study with expert users investigating the role of explainable AI (XAI) and uncertainty communication in a clinical decision support system for AI-supported nurse scheduling. Using thematic analysis of think-aloud interviews, structured questionnaires, and eye tracking data collected from 16 healthcare professionals in a real-world hospital setting, we examined how XAI and uncertainty communication affect usability, understandability, decision confidence, trust, and human–AI collaboration, with a particular HCI focus on how the design and presentation of AI outputs shape users’ attention, interpretation, reliance, and ability to use system outputs to inform concrete planning decisions.

With respect to uncertainty communication, we found that prediction interval visualizations were widely misinterpreted: participants focused on the peak or trough of the interval rather than its width as the indicator of uncertainty magnitude. Critically, adding uncertainty visualization led to a statistically significant decrease in decision confidence compared to the design without it. This suggests that exposing uncertainty alone is insufficient and may be counterproductive without carefully designed scaffolding. An actionability gap emerged as a recurring challenge: even when participants broadly understood the visualization, they struggled to translate uncertainty information into concrete planning decisions.

Regarding the explainability of AI advice, explanations were received predominantly positively and emerged as the most frequently discussed theme. Eye tracking data confirmed that explanations directed visual attention more strongly toward the AI advice itself, and participants reported greater trust and willingness to act on AI suggestions when explanations were present. However, a tension between transparency and cognitive load was evident: several participants found the explanations information-dense, and a subset did not notice them at all, pointing to a salience problem in the current design. These findings highlight that providing explanations is not sufficient in itself: their placement, timing, visual salience, and level of detail are critical interaction design choices that determine whether explanations can effectively support human decision making.

Taken together, these findings yield eight design implications: uncertainty should be communicated in simplified, example-based, and action-oriented formats; explanations should be concise, visually salient, and progressively disclosed; AI interfaces should support calibrated trust and appropriate reliance while providing explicit mechanisms for human review and override; and scheduling systems should incorporate staff preferences and fairness constraints as optimization inputs rather than post-hoc corrections. More broadly, our findings emphasize the importance of designing AI decision support as an interactive process of collaboration rather than as a one-way communication of predictions. Effective human–AI collaboration requires interfaces that preserve human agency, support critical assessment of AI outputs, and connect AI-generated information to users’ existing workflows and domain knowledge.

Future work should address several open challenges. Alternative uncertainty representations, such as frequency-based formats or hypothetical outcome plots, should be explored to better support intuitive interpretation. Additionally, future work should investigate design and interaction approaches for explaining prediction uncertainty in ways that improve both understandability and actionability. Further research should explore how AI scheduling tools can account for fairness across staff and examine how repeated interaction affects trust calibration and the risk of automation bias over time. Finally, the tension between explanation transparency and cognitive load calls for adaptive and personalized XAI interfaces that tailor explanation depth and salience to individual user needs and expertise levels. Longitudinal and in-situ evaluations would further help determine how these interaction patterns evolve with experience and how AI decision support becomes integrated into everyday work practices.

\section*{Ethics and Privacy Statement}
Ethical approval for this study was granted by the GEBHA Ethics Committee (approval number GEHBa-202601-V-434).

\section*{Acknowledgment}
This research was funded by the Bavarian State Ministry of Economic Affairs, Regional Development and Energy under grant number 41-6618c/570/2-LSM-2203-0010 as part of the KISPP research project. 
Generative artificial intelligence was used to refine grammar and style without altering the original meaning.

\section{Appendix}

\subsection{Codebook}

In the following, we describe the codebook derived from the think aloud interviews and by thematic analysis as described in the paper. We established the following 15 codes:

\begin{description}
    \item[XAI AI Advice] was used whenever the explanation about \emph{why} the rule-based AI in the AI advice design suggested a change in the shift plan was mentioned (This does not include which input the model consumes, how it was trained, or which rules it follows generally). This code concerns design variants AI Advice w/o Expl and AI Advice w/ Expl. Example: \textit{``But as I said, if you don't have to draw up the duty roster that often, the explanations make more sense.''}
    \item[Other XAI] was used whenever an explanation about the reasoning behind an AI-generated outcome (other than the one specific to code XAI AI Advice) was mentioned (This does not include the model architecture, which inputs it consumes, and how it was trained). Example: \textit{ ``Then you can add a comment explaining why more nursing staff are needed on that day.''}
    \item[Other system explanations] was used whenever a participant referred to explanatory user interface elements that help users understand the system but is unrelated to the AI model’s reasoning or decision-making process. This includes, for example, the video tutorial, color highlights, or informational texts. This code concerns all design variants. Example: \textit{Moderator:`Well, I think you actually didn't get quite enough of an explanation how to use the tool.''} Participant: \textit{``For first-time use, yes. [...]''}
    \item[System explanation of uncertainty visualization] is a specific type of system explanations and was used whenever a participant referred to the info button which explains how to interpret the visualization of predictive uncertainty. This code concerns design variants TSF w/ UV, and TSF w/ UV + PH. Example: Participant:\textit{``The lighter area and the darker area show the most likely forecast. And the width of the area indicates another [...]''}, Moderator: \textit{``Is this explanation unclear to you?''}, Participant: \textit{``It's a little difficult for me.''}
    \item[Uncertainty Communication] was coded when a participant mentioned the visualization of the prediction uncertainty. Example: \textit{``I'm currently wondering whether the uncertainty in the forecast is actually of any use to me at all.''}
    \item[Trust] was used whenever participants discussed whether they trusted the system or were willing to rely on the system. Example: \textit{``I have confidence in the system. So, probably because I'm missing some of the data right now, I'm finding there's less than there was earlier.''}
    \item[Decision Confidence] was used whenever a participant discussed how confident they felt in making a decision based on the information the system provided. Example: \textit{``So, since the program is still brand new and this is my first time checking it out, I'd say I'm still a bit unsure about my decision, of course.''}
    \item[Human-AI Collaboration] was used when participants described the interaction between people and systems, with particular emphasis on fairness, characteristics and responsibility. Example: \textit{``No, I think I'd choose a human after all at the end..''}
    \item[Experience with the System] was used whenever a participant mentioned that how they interact with the system depended on how long they are exposed to the system and how much they understand its quirks. Example: \textit{``Well, as for the uncertainty—it doesn't bother me that it's not there right now. Once I've had some experience with the system, I don't really miss it.''}
    \item[Domain Knowledge] was coded whenever a participant brought in information about personnel planning which they used to interpret the data the system presented or to make decisions. Example: \textit{``It’s worth noting that when I was writing schedules, PPR wasn’t relevant to us. I’m in an intensive care unit, and that’s something we’re still lacking here.''}
    \item[Pragmatic Quality] was used for segments in which participants describe task-oriented, functional, or usability-related aspects of user experience. Example: \textit{``Let’s be honest, as a shift planner, it’s not much use to me if the programme says that it's uncertain.''}
    \item[Hedonic Quality] was used for segments in which participants describe pleasure-related, experiential, or emotionally rewarding aspects of user experience. Example: \textit{``What amazes me is that he's highlighting this in red for the on-call shifts, but hasn't shown anything here for the overstaffing.''}
    \item[Technical Affinity] was used whenever a participant discussed their own technical abilities or made remarks about how the user interface could be perceived by others with different technical background. Example: ``[...] But since this PC thing is new to me, it's uncharted territory.''
    \item[Impact] was coded whenever the real-world consequences of using the system were explicitly mentioned (e.g., time savings, improved care quality,\ldots). Example: \textit{``If I rely on the system's recommendations or actions and they don't work out, that will naturally have negative consequences. In other words, potentially negative outcomes could arise and have consequences.''}
    \item[Requirement] was used whenever a user mentioned a feature that the current version of the system does not provide but which would be useful for creating better schedules. Example: \textit{``Because who is a caregiver? A nursing assistant has different qualifications than a certified nurse. I would definitely check that, of course. So that’s the kind of information I actually still need.''}
\end{description}

\subsection{Comparison of UEQ-S, XAI and TiA} \label{sec:apendix_questionnaires}

The following table shows the descriptive statistics for the UEQ-S~\cite{Schrepp_2017}, XAI~\cite{Hoffman_2023}, and TiA~\cite{McGrath_2025}. 

\begin{table}[htbp]
\centering
\caption{Descriptive Statistics for the Questionnaire Scores by Design}
\label{tab:descriptive_statistics_Questionnaires}

\resizebox{\textwidth}{!}{%
\begin{tabular}{llccccc}
\toprule
Questionnaire
& Scale
& Design TSF w/o UV
& Design TSF w/ UV
& Design TSF w/ UV + PH
& Design AI Advice w/o Expl
& Design AI Advice w/ Expl \\
&
& $M$ ($SD$)
& $M$ ($SD$)
& $M$ ($SD$)
& $M$ ($SD$)
& $M$ ($SD$) \\
\midrule

UEQ-S
& Pragmatic quality
& 1.56 (0.76)
& 1.12 (0.86)
& 1.25 (1.09)
& 1.98 (0.97)
& 2.06 (0.89) \\

& Hedonic quality
& 1.62 (1.00)
& 1.64 (0.75)
& 1.66 (0.86)
& 1.89 (1.13)
& 2.06 (0.85) \\

& Overall user experience
& 1.59 (0.72)
& 1.38 (0.65)
& 1.45 (0.76)
& 1.94 (1.00)
& 2.06 (0.82) \\

\addlinespace

XAI 
& XAI score
& 5.06 (0.67)
& 4.70 (0.82)
& 4.90 (0.61)
& 5.25 (0.71)
& 5.28 (0.89) \\

\addlinespace

TiA 
& TiA score
& 4.97 (0.92)
& 4.80 (0.95)
& 5.00 (1.02)
& 5.35 (1.04)
& 5.32 (0.95) \\



\bottomrule
\end{tabular}%
}

\vspace{0.5em}

\begin{minipage}{\textwidth}
\footnotesize
\textit{Note}. $n = 16$ per design condition.
$M$ = mean; $SD$ = standard deviation.
UEQ-S scores range from $-3$ to $+3$.
Higher overall TiA and TiA Trust scores indicate greater trust,
whereas higher TiA Distrust scores indicate greater distrust.
\end{minipage}
\end{table}

\bibliographystyle{unsrtnat}
\bibliography{references}

\end{document}